\documentclass[prd,article ]{revtex4}
\usepackage{graphicx}
\usepackage{amsmath}

\begin{document}

\title{Scattering solution of a ball by a bat}
\author{Alejandro Cabo$^1$, Le\'on Soler$^2$, Carlos Gonz\'alez$^3$}
\affiliation{\ $^1$ \textit{Departamento de F\'{\i}sica Te\'orica,
Instituto de Cibern\'etica, Matem\'atica y F\'{\i}sica, Calle E, No.
309, Vedado, La Habana, Cuba }}

\affiliation{$^2$ \textit{Instituto de Investigaciones del
Transporte, MITRANS, La Habana, Cuba}}

\affiliation{$^3$ \textit{Departamento de Ultras\'{o}nica, Instituto
de Cibern\'{e}tica, Matem\'{a}tica y F\'{i}sica, Calle E, No. 309,
Vedado, La Habana, Cuba}}
\begin{abstract}
\noindent The problem of the mechanical evolution  of a shock
between a cylindrically symmetric bat and a spherical ball is solved
in the strict rigid approximation for arbitrary values of the
initial conditions. The friction during the impact is assumed to
satisfy the standard rules. When the only source of energy
dissipation is friction, the problem is fully solved by determining
the separation point between the bodies. It also follows that
whatever the character of any additional form of dissipation is, it
only affects the ending value of the net impulse $I$  done by the
normal force of the bat on the ball at separation, but not the
dynamical evolution with the value of $I$ during the shock process.
A relation determining whether the contact points of the two bodies
slides between them or become at rest (to be $pure\,rotation$ state)
at the end of the impact, is determined for the case of the purely
frictional energy dissipation. The solution is also generalized  to
include losses in addition to the frictional ones and then applied
to the description of experimental measures of the scattering of  a
ball by a bat. The evaluations satisfactorily reproduce the measured
curves for the output center of mass and angular velocities of the
ball as functions of the scattering angle and the impact parameter,
respectively.

\end{abstract}

\maketitle

\newpage

\section{ Introduction}

\ Classical Mechanic is an ancient field of Physics \cite
{goldstein,landau,corben,arnold,routh,whittaker,timoshenko}. An
innumerable amount of problems had been already solved, which by now
constitute a main database for technological applications. In
particular, the scattering problem in the framework of Particle,
Nuclear and Atomic Physics has been the subject of intense
investigation along centuries of research \cite {mott,newton,wu}. On
another hand, as stated in Ref. \cite{nathan}, at difference with
the situation in microscopic Physics, the scattering between
macroscopic bodies, had not been a similarly attended area of study.
However, in relatively recent times, and as motivated by the
relevance within the baseball of the shocking process between the
bat and the ball in the baseball sport, a research activity on the
theme had been stimulated \cite{stronge, stronge1,nathan, nathan1}.
An extended study of the physical process in impact mechanics can be
found in Ref. \cite{stronge}. References \cite{stronge1, nathan}
presented detailed studies of the shock problem of a bat and a ball
directed to investigate the optimal batting configurations and the
scattering results of the impact process at low velocities. The
solution of the shock process given in those works were found under
the restrictions of: 1) an assumed two dimensional character of the
shock, and 2) the use of particular models for the impulse forces
appearing during the impact. The work presented in Ref. \cite{
nathan1} is devoted to investigate the influences of the lack of
rigidity of the bat and the ball, and the dynamical evolution of the
ball in the air on the bating process results. In these cited works
references to a number of additional studies stimulated by the
relevance of mechanical processes in sports can be found.

In the present study we investigate the  problem of the shock
between a rigid and spherical ball and an also rigid bat showing a
cylindrical symmetry axis. one of the  aims is to generalize the
results of \ Ref. \cite{stronge1, nathan} for arbitrary initial
conditions for the center of mass and angular velocities. In
addition we here apply the results to the description of the
experimental measures of the scattering of a ball by a bat presented
in Ref. \cite{nathan}. For bookkeeping purposes, the simpler
explicit solution of the conservative and friction less impact is
also presented in an appendix.

\ The discussion starts by considering the case in which sliding
frictional forces in the contact points develops in the assumed
strictly rigid bat and ball. As mentioned above, the situation
generalizes the one studied in Refs. \cite{stronge1,nathan} , by
removing their two main assumptions: 1) The shock will considered
asfully three dimensional with arbitrary initial conditions, and the
bat form is only restricted to be a solid of rotation showing a
cylindrical symmetry axis, 2) No model about the nature of the
normal impact force will be adopted. Only the standard connection
between the sliding friction and the normal force will be assumed.
That is, the modulus of the sliding friction force will be equal to
the friction coefficient $\mu $ times the magnitude of the
instantaneous normal force at the contact point. As usual, the
friction force over one of the bodies will be directed in the
opposite sense to the tangent component for the relative velocity of
the contact point on that object with respect to the contact point
on the other body. \

In addition, a  criterium is found for deciding about whether the
ending state of the shock corresponds to sliding tangent surfaces, or to a $%
pure$ $rotation$ state.  The $pure$ $rotation$ state will be called
that one in which the tangent surfaces end the shocking process by
showing null relative velocity. \ Analytic and integral expressions
of the ending values of the center of mass and angular velocities
are given, in terms of the solution of a simple differential
equation for the two components of the relative velocities between
the tangent points. It is an interesting outcome that the impulse
done by the normal  force of the bat on the ball can be employed in
place of the time in describing the dynamical evolution during the
shock. \

\ The solution for the ending velocities of the ball and the bat
presents two kinds of behavior: In one case, the friction force is
unable to reduce the tangential component of the relative velocity
to zero during the small time interval of the shock. In this option,
the two bodies end the impact with a remaining sliding between their
contact surfaces. \

In the alternative ending state, the frictional force becomes able
in reducing to zero the relative velocity of the two contact points
between the bodies at an intermediate instant of the shocking
interval. In general, at this time instant $t_{rp}$ in which the
sliding between surfaces ends, the normal force is not yet
vanishing. This means that a finite portion of the initial
mechanical energy of the system is yet stored in the form of
$elastic$ deformation energy. Therefore, after the attaining of the
$pure$ $rotation$ state, the system, which is yet within the
shocking  process, evolves in an alternative way than in the
previous $sliding$ period. In this second interval, the evolution
becomes conservative, and the equations are similar but not
identical to the ones corresponding to the friction less
conservative problem solved in the Appendix A. Their difference
rests in that within this period, a static kind of frictional force
can contribute to the conservative mutual impulses between the
bodies.

In order to  make the solution applicable to the realistic cases (in
which there exist appreciable energy losses due heat, deformations,
sound, etc., in addition to the frictional one), the solution of the
problem allowing only frictional dissipation is here generalized to
include alternative energy losses.  The generalization was suggested
and helped  by two factors:  a)  The possibility  of properly
identifying the amount of  stored elastic energy in the system at
any moment when the energy losses are purely due to friction; b) The
helpful technical fact  that due to the assumed extreme rigidity
approximation, the exact evolution of the system, no matter the
nature of the energy losses, only depends on one single variable:
the net impulse $I(t) $  transmitted up to a given instant $t$ by
the normal force exerted by the bat on the ball.

The generalized analysis is then applied  to the description of the
 experiments on the scattering of a ball by a bat  reported in
Ref. \cite {nathan}. \ After phenomenologically fixing a single
experimentally measured quantity: the value of the center of mass
velocity of the ball at zero values of the impact parameter and its
initial angular velocity, the calculated  results furnish a
satisfactory description of the reported  measurements.  In
particular, the curves for the ending velocity  of the ball  as
functions of the scattering  angle coincided with the  measured ones
within  the range of the dispersion of the data, for each of the
three values of the initial angular velocity of the ball employed in
the experiences.

The exposition of the work proceeds as follows. In section II, the
equations for the shock problem are presented and the notation and
basic definitions are given. In Section II the solution of the
impact problem for the case in which the ending state is assumed to
corresponds to sliding contact surfaces is exposed.  The Section IV
continues  by presenting the solution for the situation in which the
contact point of the ball and the bat finish the shock process in
the $pure$ $rotation$ state.  Finally, Section V the results of
previous sections are  applied to solve the scattering problem of
the ball by the bat in the  particular  configuration considered in
the experiments reported in Ref. \cite{nathan}. The results for the
description of the measurement reported in that reference are
presented. Finally, Section V a summary of the results is exposed.

\section{Energy dissipative impact: the sliding final state}

\ This section will expose some basic considerations and definitions
which will be of use along the presentation.  The figure 1,
illustrates the shock process between the ball and the bat in the
precise instant at which they become in contact. The adopted
laboratory system of reference (to be named as the $Lab$ system in
what follows), will be situated on the center of mass of the bat and
having its $z$$(x_3)$ coordinate axis being collinear with the
cylindrical symmetry axis of the bat. The three unitary vectors of
the $Lab$ reference frame axes will be $\{\mathbf{i,j,k}\}$. In what
follows, bold letters will indicate vectors. The unit vector
$\mathbf{k}$ along the $z$ axis, will point in the direction of the
barrel of the bat, and thus, the vectors $ \mathbf{i,j}$ will be
contained in a transversal section of the bat as illustrated in
figure 1, and they are chosen to form a direct triad with
$\mathbf{k}$. \ The vector $\mathbf{r}_c$ depicted in figure 1,
defines the position of the contact point of the two bodies in the
above defined reference frame.  Note that the adoption of the $Lab$
frame does not restrict the generality of the discussion. The vector
$\mathbf{r}_{cmp}$ gives the position of the center of mass of the
ball in the $Lab$ coordinate
system.  The set of three unit vectors $\{\mathbf{t}_1\mathbf{,t}_2\mathbf{%
,t}_3\}$ are defined as follows: $\mathbf{t}_3$ is normal to the
common tangent surface of both bodies at the contact point, and is
directed as pointing outside the volume of the bat. Further,
$\mathbf{t}_2$ can be defined as a tangential unit vector being
contained in a common plane with the unit vector $\mathbf{k}$ and
having a positive scalar product with it. Finally, $\mathbf{t}_1$ is
defined as being orthogonal to $\mathbf{t}_2$ and
$\mathbf{t}_3$\thinspace by also forming with them a direct triad.
\begin{figure}[h]
\hspace*{-0.4cm} \includegraphics[width=8.5cm]{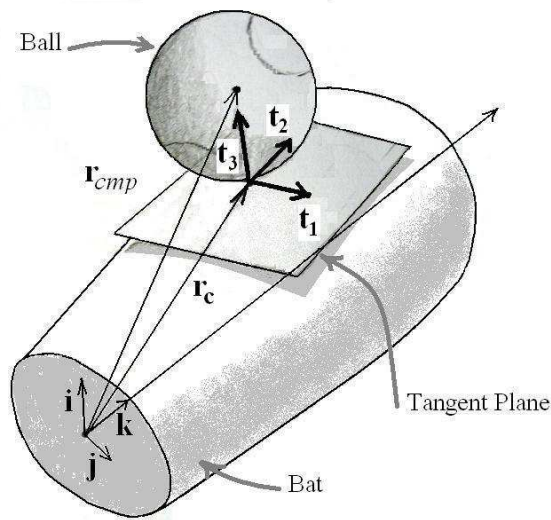}
\caption{ The figure illustrate the defined systems of reference to
be systematically employed along the work. The ball and the bat are
shown in the moment at which the impact starts. } \label{multi}
\end{figure}

\ Let us specify  now few physical assumptions that will be adopted
for the solution of the problem.  Firstly, as it was already stated,
it will be considered that the ball and the bat are ideally rigid
bodies. That is, the elastic forces are supposed as being so strong,
that the spacial forms of
both objects remain almost the same during the whole shocking time interval $%
\delta t_{ch}.$ Naturally, this lapse will be supposed as being extremely
short. These assumptions will be reflected in the discussion to follow, in
which the whole geometry of the arrangement will be supposed to be
invariable during the time interval $\delta t_{ch}.$ \ The only quantities
that will allowed to change are the linear and angular velocities of the two
bodies.

\ Let us write the general equations of motions for the evolution of the
center of mass linear momenta $\mathbf{P}_p$, $\mathbf{P}_b$ and the angular
momenta $\mathbf{L}_p,\mathbf{L}_b$ of the ball and the bat, respectively,
during the shocking interval $\delta t_{ch}$. They can be written as
\begin{eqnarray}
\text{\ }\frac d{dt}\mathbf{P}_p(t) &=&\mathbf{F}(t),\text{ \ } \\
\frac d{dt}\mathbf{P}_b(t) &=&-\mathbf{F}(t),\text{ \ \  } \\
\text{\ }\frac d{dt}\mathbf{L}_p(t) &=&(\mathbf{r}_c-\mathbf{r}_{cmb})\times
\mathbf{F}(t),\text{ }  \label{Lball} \\
\text{\ }\frac d{dt}\mathbf{L}_b(t) &=&-\mathbf{r}_c\times \mathbf{F}(t),%
\text{ \ }
\end{eqnarray}
in which $\mathbf{F}(t)$ is the very rapidly varying impulsive force which
is exerted by the bat on the ball. \ The definitions of the momenta are
\begin{eqnarray}
\mathbf{P}_b(t) &=&m_b\mathbf{v}_{cmb}=m_b\frac d{dt}\mathbf{r}_{cmb}(t),
\nonumber \\
\mathbf{P}_p(t) &=&m_p\mathbf{v}_{cmp}=m_p\frac d{dt}\mathbf{r}_{cmp}(t),
\nonumber \\
\mathbf{L}_b(t\text{ }) &=&\widehat{\mathbf{I}}_b\cdot \mathbf{w}_b,
\nonumber \\
\mathbf{L}_p(t) &=&\widehat{\mathbf{I}}_p\cdot \mathbf{w}_p(t)+\mathbf{r}%
_{cmp}(t)\times m_p\stackrel{\mathbf{.}}{\mathbf{v}}_{cmp}(t),
\label{angdef}
\end{eqnarray}
in which $m_b$ and $m_p$ are the masses of the bate and the ball,
respectively. The angular momenta of the bat is defined with respect to the $%
Lab$ reference frame, and for the simplicity of the further
discussion  the angular momenta of the ball was defined  with
respect  to its center of mass. \ Note that the angular impulse in
equation (\ref{Lball}) is consistent with this definition. The
inertia tensor of the ball $\widehat{\mathbf{I}}_p$ is the identity
matrix and the one associated to the bat $ \widehat{\mathbf{I}}_b$\
has cylindrical symmetry  in the $Lab$ reference system. The
explicit forms of the inertia tensor are
\begin{equation}
\widehat{\mathbf{I}}_p=I_p\left(
\begin{array}{lll}
1 & 0 & 0 \\
0 & 1 & 0 \\
0 & 0 & 1
\end{array}
\right) ,\text{ \ \ }\widehat{\mathbf{I}}_b=\left(
\begin{array}{lll}
I_1 & 0 & 0 \\
0 & I_2 & 0 \\
0 & 0 & I_3
\end{array}
\right) =\left(
\begin{array}{lll}
I_t & 0 & 0 \\
0 & I_t & 0 \\
0 & 0 & I_3
\end{array}
\right) .
\end{equation}

Note, that the vector $\mathbf{r}_{cmb}(t)$ is the position of the
center of mass of the bat, which during the shocking interval
$\delta t_{ch}$ remains to be very close to the origin of the $Lab$
reference frame depicted in figure 1, if the bodies are sufficiently
rigid for the given initial relative velocities and angular momenta.
That is, in the here adopted strict rigidity approximation, in which
the two bodies are assumed be completely invariable
in form and position during the impact, it will assumed that $\mathbf{r}%
_{cmb}(t)=0.$

In a first instance, we will include energy dissipation only through
the presence of sliding friction during the impact. The inclusion of
additional sources of energy losses will be incorporated in last
sections . As stated in the introduction the usual laws of friction
will be assumed. That is, we will consider that the modulus of the
sliding friction vector will be the sliding friction coefficient
$\mu $, times the modulus of the normal force between the bodies.
Further, the direction of the friction force over one of the two
bodies contact point, will be opposite to the relative velocity of
this point with respect to the contact point of the other body.

 Let us note that, because the normal force grows starting from zero at the
beginning of the shocking period, in general the first stage of the shock
process should correspond to a situation in which the contact points of the
ball and the bat slide between them at the beginning of the process. Only in
the particular situation in which \ the tangent velocity is already
vanishing at the beginning of the impact, this period will not exist. In
such a case the solution is directly given by the \ one presented in the
next section.

Then, consider the equations of motion as written for this initial
process. For an instant $t$  being inside the very small time
interval $\delta t_{ch} $ during which the bodies are in contact,
the equations can be written in the form
\begin{eqnarray}
m_p\,d\mathbf{v}_{cmp}(t) &=&(-\mu \frac{\mathbf{v}_{pb}^{(t)}}{\mid \mathbf{%
v}_{pb}^{(t)}\mid }+\mathbf{t}_3)\text{ }dI\ , \label{NewtonIN}\\
m_b\,d\mathbf{v}_{cmb}(t) &=&(\mu \frac{\mathbf{v}_{pb}^{(t)}}{\mid \mathbf{v%
}_{pb}^{(t)}\mid }-\mathbf{t}_3)\text{ }dI\ ,\text{ } \\
I_p\text{ }d\mathbf{w}_p(t) &=&-\mu \text{\ }dI\text{ }(\mathbf{r}_c-\mathbf{%
r}_{cmp})\times \frac{\mathbf{v}_{pb}^{(t)}}{\mid \mathbf{v}_{pb}^{(t)}\mid }%
, \\
d\,\mathbf{w}_b(t) &=&\mu \text{ }dI\text{ \ }\widehat{\mathbf{I}}%
_b^{-1}\cdot \mathbf{r}_c\times \frac{\mathbf{v}_{pb}^{(t)}}{\mid \mathbf{v}%
_{pb}^{(t)}\mid }-\frac 1{I_t}\text{\ }dI\text{ }\mathbf{r}_c\times \mathbf{t%
}_3,  \label{Newton2}
\end{eqnarray}
where a new magnitude appearing is the differential impulse $%
dI=N_{b\rightarrow p}dt$ done by the normal force $N_{b\rightarrow p}$
exerted by the ball on the bat. It allows to define the net impulse done by
this force up to a given time $t$ by
\begin{equation}
I(t)=\int_0^tN_{b\rightarrow p}dt.  \label{Nimpulse}
\end{equation}
Another quantity appearing is the tangential component of the relative
velocity $\mathbf{v}_{pb}^{(t)}$ between the contact point of the ball and
the corresponding contact point of the bat (for which $\mid \mathbf{v}%
_{pb}^{(t)}\mid $ means its modulus). \ \ As defined above, $\mu $ is the
sliding friction coefficient and the inverse of the inertia moment tensor of
the bat $\widehat{\mathbf{I}}_p^{-1}$ can be explicitly written as follows
\begin{eqnarray}
\widehat{\mathbf{I}}_b^{-1} &=&\left(
\begin{array}{lll}
\frac 1I & 0 & 0 \\
0 & \frac 1I & 0 \\
0 & 0 & \frac 1{I_3}
\end{array}
\right) =\left(
\begin{array}{lll}
\frac 1I & 0 & 0 \\
0 & \frac 1I & 0 \\
0 & 0 & \frac 1I
\end{array}
\right) +\left(
\begin{array}{lll}
0 & 0 & 0 \\
0 & 0 & 0 \\
0 & 0 & \frac 1{I_3}-\frac 1I
\end{array}
\right)   \nonumber \\
&=&\frac 1I\mathbf{\delta }+(\frac 1{I_3}-\frac 1I)\,\mathbf{k}\,\mathbf{k,}
\nonumber \\
\mathbf{i} &=&(1,0,0),\text{ }\mathbf{j}=(0,1,0)\text{ and }\mathbf{k}%
=(0,0,1),
\end{eqnarray}
where $\mathbf{k}\,\mathbf{k}$ means the diadic tensor $\mathbf{k}$ $\mathbf{%
k\equiv }k_i\,k_j.$

\ Let us define now a simplified notation for the tangential component $%
\mathbf{v}_{pb}^{(t)}(t)$ of the  full relative velocity $\mathbf{v}_{pb}(t)$%
\ between  the contact points of the ball  the bat as follows
\begin{eqnarray}
\mathbf{v}_{pb}^{(t)}(t) &\equiv &\mathbf{v}(t)=v_1(t)\text{ }\mathbf{t}%
_1+v_2(t)\text{ }\mathbf{t}_2,  \nonumber \\
& &\mid \mathbf{v}\mid \equiv v(t)=\sqrt{(v_1)^2+(v_1)^2.}
\end{eqnarray}

Then, $\mathbf{v}_{pb}(t)$ can be written as follows
\begin{eqnarray}
\mathbf{v}_{pb}(t) &=&\mathbf{v}_{pb}^{(t)}+(\mathbf{v}_{pb}(t)\cdot \mathbf{%
t}_3)\text{ }\mathbf{t}_3  \nonumber \\
&=&v_1(t)\text{ }\mathbf{t}_1+v_2(t)\text{ }\mathbf{t}_2+(\mathbf{v}%
_{pb})\cdot \mathbf{t}_3\text{ }\mathbf{t}_3  \nonumber \\
&=&\mathbf{v}_{cmp}(t)-\mathbf{v}_{cmb}(t)+\mathbf{w}_p(t)\times (\mathbf{r}%
_c-\mathbf{r}_{cmp})-\mathbf{w}_b(t)\times \mathbf{r}_c,  \label{vtangente}
\end{eqnarray}
which allows to write for the variation of $\mathbf{v}_{pb}(t)$ in a time
interval $dt$ in the considering sliding interval the expression
\[
d\text{ }\mathbf{v}_{pb}(t)=d\text{ }\mathbf{v}_{cmp}(t)-d\text{ }\mathbf{v}%
_{cmb}(t)+d\text{ }(\mathbf{w}_p(t))\times (\mathbf{r}_c-\mathbf{r}_{cmp})-d%
\text{ }(\mathbf{w}_b(t))\times \mathbf{r}_c,
\]
where, as described before, the geometry of the system has been
assumed as invariant due to the perfect rigidity of the ball and the
bat. \ \ This assumption can be satisfied, in particular, if all the
velocities are chosen to be scaled to sufficiently small values
producing small \ enough deformations. Employing the equations of
motion allows to express the above variation in terms of the
differential impulse of the normal forces in the following linear
form
\begin{eqnarray}
d\text{ }\mathbf{v}_{pb}(t)=-\mu (\frac 1{m_p}+\frac 1{m_b})\frac{\mathbf{v}%
(t)}{v(t)}dI &&+(\frac 1{m_p}+\frac 1{m_b})\mathbf{t}_3\text{ }dI  \nonumber
\\
\text{ \ }-\frac \mu {I_p}dI\text{ }(\mathbf{r}_c-\mathbf{r}_{cmp})\times
{\Large (}(\mathbf{r}_c-\mathbf{r}_{cmp})\times \frac{\mathbf{v}(t)}{v(t)}%
{\Large )} &&+\mu \text{ }dI\text{ \ }\widehat{\mathbf{I}}_b^{-1}\cdot
\mathbf{r}_c\times \frac{\mathbf{v}(t)}{v(t)}  \nonumber \\
&&-\frac 1{I_t}\text{\ }dI\text{ }\mathbf{r}_c\times \mathbf{t}_3.
\end{eqnarray}

After projecting the above relation over $\mathbf{t}_1$ and $\mathbf{t}_2$ ,
the following set of two differential equations for the variation of the
relative tangential velocity components as functions of the impulse of the
normal force of the bat on the ball can be obtained
\begin{eqnarray}
dv_1(I) &=&-s_1\frac{v_1(I)}{\sqrt{(v_1(I))^2+(v_2(I))^2}\,}dI,  \nonumber \\
dv_2(I) &=&-s_2\frac{v_2(I)}{\sqrt{(v_1(I))^2+(v_2(I))^2}}\text{ }dI+s_0%
\text{ }dI\text{\ ,\ \ \ \ }  \label{vtan}
\end{eqnarray}
in which the parameters appearing depend on the system's properties
as follows
\begin{eqnarray}
s_1 &=&\mu \left( \frac 1{m_p}+\frac 1{m_b}+\frac{(\mathbf{r}_c-\mathbf{r}%
_{cmp})^2}{I_p}+\frac{(\mathbf{r}_c)^2}{I_t}+(\frac 1{I_3}-\frac 1{I_t})%
\text{ }(\mathbf{t}_1\cdot (\mathbf{r}_c\times \mathbf{k}))^2\right) \text{\
\  },  \label{s1} \\
s_2 &=&\mu \left( \frac 1{m_p}+\frac 1{m_b}+\frac{(\mathbf{r}_c-\mathbf{r}%
_{cmp})^2}{I_p}+\frac{(\mathbf{r}_c)^2}{I_t}-\frac{(\mathbf{r}_c\cdot
\mathbf{t}_2)^2}{I_t}\right) ,  \label{s2} \\
s_0 &=&(\mathbf{r}_c\cdot \mathbf{t}_3)(\mathbf{r}_c\cdot \mathbf{t}_2).
\label{s0}
\end{eqnarray}

At this point it can be underlined that the time had disappeared
form the equations of motion for the tangent velocities. That is,
the time dependence is all embodied  in the time dependence of  net
impulse of the normal force of the bat on the ball  $I(t)$ defined
 by (\ref{Nimpulse}). This  functional relation between the
impulse $I$ and the time $t$ will be assumed in what follows  for
studying the the evolution of the system during the shock interval
in terms the impulse $I$ in place of the time $t$. \ A very
important consequence of the unique dependence of the evolution on
the net impulse $I$, is the fact the presence of any form of
dissipation in addition to friction will not affect the result of
the evolution of all the properties up to a given state with
definite value of $I$. Therefore, the only effect of the presence of
such supplementary forms of dissipation  will be to determine a
different  instants in which separation (end of the impact) will
occur.  This property will be employed in last sections to  extend
the solution  to include non dissipation effects.

The solution of the above set of equations for the relative
velocities, after determining their initial values allows to
determine the evolution with the impulse $I$ of all these
velocities. As it will be verified in the particular examples solved
in next sections, the general behavior of the solutions is such that
both components evolve with $I$ in a continuous  way up to a
critical value of the normal impulse, at which  both components
simultaneously tend to vanish. This point correspond to the
attainment  of the  $pure$ $rotation$ state. However, whether this
critical situation would be approached or not  depends on the
dynamical equations of motion: they in fact should determine whether
or not  positive values of the normal force done by  the bat on the
ball  can exist up to the arriving to the $pure$ $rotation$ state.
Verifying the above remarks,  the set of equations of motion
(\ref{NewtonIN}-\ref{Newton2}) \ during the impact process can be
written in terms of the most appropriate evolution parameter $I$ as
follows
\begin{eqnarray}
m_p\,\frac{d\mathbf{v}_{cmp}(I)}{dI} &=&(-\mu \frac{\mathbf{v}(I)}{v(I)}+%
\mathbf{t}_3) ,  \label{vp(I)} \\
m_b\,\frac{d\mathbf{v}_{cmb}(I)}{dI} &=&(\mu \frac{\mathbf{v}(I)}{v(I)}-%
\mathbf{t}_3) ,  \label{vb(I)} \\
I_p\text{ }\frac{d\mathbf{w}_p(I)}{dI} &=&-\mu \text{\ }(\mathbf{r}_c-%
\mathbf{r}_{cmp})\times \frac{\mathbf{v}(I)}{v(I)},  \label{wp(I)} \\
\frac{d\,\mathbf{w}_b(I)}{dI\text{ }} &=&\mu \text{ \ }\widehat{\mathbf{I}}%
_b^{-1}\cdot \mathbf{r}_c\times \frac{\mathbf{v}(I)}{v(I)}-\frac 1{I_t}\text{%
\ }\mathbf{r}_c\times \mathbf{t}_3.  \label{wb(I)}
\end{eqnarray}

These equation clearly evidence that the evaluation of all the velocities is
only determined by the value of the \ impulse of the normal force $I$ \
being exerted by the bat on the ball. This impulse is defined by equation (%
\ref{Nimpulse}). \ Since the solution of the differential equations
(\ref{vtan})) determines the tangential velocity as a function of
$I$ , the equations (\ref{vp(I)})- (\ref{wb(I)}) can be \ integrated
\ to find out all the velocities in terms of $\ I$ \ as follows
\begin{eqnarray}
\,\mathbf{v}_{cmp}(I) &=&\mathbf{v}_{cmp}(0)+\frac 1{m_p}(-\mathbf{I}%
_{fr}(I)+I\text{ }\mathbf{t}_3)\text{, }  \label{vpsol} \\
\,\mathbf{v}_{cmb}(I) &=&\mathbf{v}_{cmb}(0)+\frac 1{m_b}(\mathbf{I}%
_{fr}(I)-I\text{ }\mathbf{t}_3)\text{ },
\label{vbsol} \\
\mathbf{w}_p(I) &=&\mathbf{w}_p(0)-\frac 1{I_p}\text{ }(\mathbf{r}_c-\mathbf{%
r}_{cmp})\times \mathbf{I}_{fr}(I),  \label{wpsol} \\
\mathbf{w}_b(I) &=&\mathbf{w}_b(0)\text{ }+\text{\ }\widehat{\mathbf{I}}%
_b^{-1}\cdot (\mathbf{r}_c\times \mathbf{I}_{fr}(I))-\frac 1{I_t}\text{\ }I%
\text{ }\mathbf{r}_c\times \mathbf{t}_3,  \label{wbsol}
\end{eqnarray}
where the impulse done by the tangential frictional force is defined
as a function of $I$ by \ the following integral
\begin{equation}
\mathbf{I}_{fr}(I)=\mu \int_0^IdI^{\prime }\text{ }\frac{\mathbf{v}%
(I^{\prime })}{v(I^{\prime })}.  \label{frictionImpulse}
\end{equation}

For a coming reference to them, let us define now the increments in
the velocities with respect to their
initial values $\mathbf{v}_{cmp}^{(in)},\mathbf{v}_{cmb}^{(in)},\mathbf{w}%
_p^{(in)}$ and $\mathbf{w}_b^{(in)}$when \ the  impulse rises to a definite
value $I(t)$ by:
\begin{eqnarray}
\Delta \text{ }\mathbf{v}_{cmp}(I) &=&\mathbf{v}_{cmp}(I)-\mathbf{v}%
_{cmp}(0),  \nonumber \\
\Delta \text{ }\mathbf{v}_{cmb}(I) &=&\mathbf{v}_{cmb}(I)-\mathbf{v}%
_{cmb}(0),  \nonumber \\
\Delta \mathbf{w}_p(I) &=&\mathbf{w}_p(I)-\mathbf{w}_p(0),  \nonumber \\
\Delta \mathbf{w}_b(I) &=&\mathbf{w}_b(I)-\mathbf{w}_b(0),  \label{increm} \\
\mathbf{v}_{cmp}(0) &=&\mathbf{v}_{cmp}^{(in)},\text{ }\mathbf{v}_{cmb}(0)=%
\mathbf{v}_{cmb}^{(in)},  \nonumber \\
\mathbf{w}_p(0) &=&\mathbf{w}_p^{(in)},\text{ }\mathbf{w}_b(0)=\mathbf{w}%
_b^{(in)}.  \label{init}
\end{eqnarray}

 The  above formulae indicate that the shock problem in this assumed \
initial sliding process, will become solved, after finding a condition
determining the final value of the   tangent velocity component at the
moment of separation of the bodies. \ Let us consider this point in what
follows.

 In the assumed in this section case, in which the shock is finalized when the
contact points are yet sliding between them, the posed equations
remain being valid along all the shock interval.  In this situation,
the appropriate condition for fixing the ending value of the impulse
$I$ is that total dissipative work $W$ done (due to friction or non
elastic processes) up to the value of the impulse $I$, should be
equal to the decrease of the total kinetic energy $\Delta E_{kin}$\
along the evolution up to the same value of  $I$ . \ The physical
reason for this condition is that when the normal force vanishes,
which defines the separation of the bat and the ball, the
conservation of energy implies that all the non already dissipated
part of the mechanical energy should appear in the form of the
translational and rotational kinetic energies of the bat and the
ball.

Therefore, since as it  has been concluded,  the evolution as a function of $%
I$ is completely independent of the nature of the dissipation, it
follows that the solution of the problem in this period is only
depending of the fraction of the initial  kinetic  energy of the two
bodies which becomes dissipated  in the shocking process, a quantity
which only will determine the value of the impulse transmitted at
separation $I_{out}$.
This condition leads to the equation for $I_{out}$%
\begin{equation}
W(I_{out})=\Delta E_{kin}(I_{out}). \label{cond}
\end{equation}

The part of the total dissipative work $W$ which is done by the
friction up to the value of the time $t$, for which the impulse has
the value  $I(t)$ can be calculated to be
\begin{align}
W_{fr}(I(t))& =-\int_0^t\mu N_{b\rightarrow p}\mid \mathbf{v}\mid dt
\nonumber \\
& =-\mu \int_0^{I(t)}\mid \mathbf{v(}I\mathbf{)}\mid dI.  \label{wfriction}
\end{align}

In what follows, \ we will also consider also the existence of \ additional
sources of dissipation, as the one associated to the non complete elastic
behavior of the ball and the bat. Then, the total dissipative work $W(I)$
will be written as
\begin{equation}
W(I)=W_{fr}(I)+W_{add}(I),
\end{equation}
where $W_{add}(I)$ represents  the amount of mechanical energy
dissipated in the system  up to the instant in which the normal
impulse takes the value $I$ due to mechanisms additional to the
frictional one. It can be understood that the dependence  on $I$ of
$W_{add}$ $(I)$ will depend on the \ concrete forms of the
dissipation process acting in the bat at the ball.  However, it is a
remarkable property that, assumed that if the system becomes able to
arrive to the pure rotation state, the velocities at this point
result to be (in the here considered perfect rigidity situation and
only at this particular instant) completely independent of the
existence  of non frictional kinds of dissipation. \ The particular
cases of the experimental results to be considered in next sections
belong to the situation in which $pure$ $rotation$ is established.

The explicit form of the condition (\ref{cond}) determining the
separation point\ under sliding regime (in case that it effectively
occurs) is completed after defining the formula for the increase in
the total kinetic energy as a function of $I.$  It follows after
expressing the values of the final lineal and angular velocities in
terms of their initial values plus their increment. Making use of
the equations (\ref{increm}) this quantity has the expression
\begin{eqnarray}
\Delta E_{kin}(I) &=&\frac{m_p}2\left( 2\mathbf{v}_{cmp}^{(in)}\cdot \Delta
\mathbf{v}_{cmp}(I)+(\Delta \mathbf{v}_{cmp}(I))^2\right) +  \nonumber \\
&&\frac{m_b}2\left( 2\mathbf{v}_{cmb}^{(in)}\cdot \Delta \mathbf{v}%
_{cmb}(I)+(\Delta \mathbf{v}_{cmb}(I))^2\right) +  \nonumber \\
&&\frac{I_p}2\left( 2\mathbf{w}_p^{(in)}\cdot \Delta \mathbf{w}_p(I)+(\Delta
\mathbf{w}_p(I))^2\right) +  \nonumber \\
&&\frac 12\left( 2\mathbf{w}_b^{(in)}\cdot \mathbf{I}_b\mathbf{\cdot .}%
\Delta \mathbf{w}_b(I)+\Delta \mathbf{w}_b(I)\cdot \mathbf{I}_b\cdot \Delta
\mathbf{w}_b(I)\right) ,
\end{eqnarray}
in which all the increments in the linear and angular velocities are
given by the formulae furnished by the integrated equations of
motion (\ref{vpsol}-\ref{wbsol}) and the solution of the
differential equations for the tangent velocity (\ref{vtan}).

\ Therefore, the full solution of the problem in the assumed situation  in
which along all  the shock process the contact points of the ball and bat
slide between them, can be obtained after finding the value of $I_{out}$
making equal to zero the function
\begin{equation}
S(I_{out})=W(I_{out})-\Delta E_{kin}(I_{out}).  \label{decision2}
\end{equation}

It should be remarked that for the case in which all the dissipation
is determined by friction, the problem is completely solved because
the evaluation of the work of friction as a  function of $I_{out}$
is completely defined by (\ref{wfriction}) in terms of the found
solutions of the tangential velocities as functions of the impulse
$I.$ \ \ The case of additional sources of losses, for to be
analogously solved needs for a definition of the specific mechanism
leading to dissipation. \ \ A model for taking in consideration such
a mechanism will be constructed in the coming section IV in order to
apply the results to the description of realistic experimental
measures given in Ref.\cite{nathan}.

This completes the finding of the solution for the evolution of the
impact process  in this firstly assumed situation. It can be helpful
to recall that all the shocking events in which the contact points
slide at the beginning of the impact, start evolving as guided by
this just  considered  $sliding$ case. In the case that   no
solution exists for the above equation, the situation should
correspond to a problem in which  the work which is done by the
friction along the  whole  initial sliding period is not able to
become equal to the decrease in the kinetic energy of the bat and
the ball.  Let us consider the finding of the solution  in this
second regime.

\section{Energy dissipative impact: the ''pure rotation'' final state}

\ As followed from the previous section, we will now consider  the
situation in which the maximal work that can be done by the friction
up to the point in which the sliding between the contact points
vanish, is not able to dissipate  the mechanical energy down to the
value  needed to coincide with the total kinetic  energy at this
same sliding state. In other words, at the
instant $t_{rp}$ (or$\ $the corresponding impulse of the bat on the ball $%
I_{rp})$ at which sliding stops, and the $pure$ $rotation$ state is
attained, a portion of the total initial energy should be yet stored
in the form of  deformation energy. In addition,   the presence of
deformation energies forces is represented in the considered problem
by a non vanishing normal force.

Thus,  the shock process is divided in two parts, each one being
governed by different dynamical equations. \ The first one was
discussed in the past section, in which  sliding occurs and  stops
at the impulse value $I_{rp}$ at which the slice between the contact
points of the ball and bat ends. At this value of the impulse  the
increments in the velocities are given by the formulae
(\ref{vpsol}-\ref{wbsol}), in which the tangential sliding velocity
vanishes. \ These linear and angular velocity increments have the
explicit expressions
\begin{eqnarray}
\,\mathbf{v}_{cmp}(I_{rp}) &=&\mathbf{v}_{cmp}(0)+\frac 1{m_p}(-\mathbf{I}%
_{fr}(I_{rp})+I_{rp}\text{ }\mathbf{t}_3),
\label{vpPR} \\
\,\mathbf{v}_{cmb}(I_{rp}) &=&\mathbf{v}_{cmb}(0)+\frac 1{m_b}(\mathbf{I}%
_{fr}(I_{rp})-I_{rp}\mathbf{t}_3),  \label{vbPR} \\
\mathbf{w}_p(I_{rp}) &=&\mathbf{w}_p(0)-\frac 1{I_p}\text{ }(\mathbf{r}_c-%
\mathbf{r}_{cmp})\times \mathbf{I}_{fr}(I_{rp}),  \label{wpPR} \\
\mathbf{w}_b(I_{rp}) &=&\mathbf{w}_b(0)\text{ }+\text{\ }\widehat{\mathbf{I}}%
_b^{-1}\cdot (\mathbf{r}_c\times \mathbf{I}_{fr}(I_{rp}))-\frac 1{I_t}\text{%
\ }I_{rp}\text{ }\mathbf{r}_c\times \mathbf{t}_3,  \label{wbPR}
\end{eqnarray}

These quantities fully determine the linear and angular velocities
of the ball and the bat in the next intermediate  $pure$ $rotation$
state in which a portion of the total mechanical energy is yet
stored in the form of elastic
deformation energy. The value of the total kinetic energy at this point $%
E^{(rp)}$ can be calculated through the formula
\begin{eqnarray}
E^{(rp)} &=&\frac{m_p}2(\mathbf{v}_{cmp}(I_{rp}))^2+\frac{m_b}2(\mathbf{v}%
_{cmb}(I_{rp}))^2+  \nonumber \\
&&\frac{I_p}2(\mathbf{w}_p(I_{rp}))^2+\frac 12\mathbf{w}_b(I_{rp})\cdot
\mathbf{I}_b\cdot \mathbf{w}_b(I_{rp}).  \label{Erp}
\end{eqnarray}

As remarked before, the existence of  deformation energy at the
transmitted impulse $I_{rp}$ is represented by the existence of a
yet not vanishing normal force, which should tend to zero in the new
stage of evolution. In this second period of the impact process, the
stored elastic energy transforms in a contribution to the kinetic
energy of the ball and the bat at the real ending state of the
shock, in which the normal force tends to vanish. Deformation losses
can occurs also in this process.

Therefore, after the instant at which the impulse is $I_{rp}$ , the
system will be governed by a similar, but not identical, set of
equations valid for the conservative shock studied in appendix A.
The difference is related with the fact that during this last
interval, a $static$ frictional force can be dynamically required to
remain acting. This possibility was excluded in the case of the
absence of friction of the conservative case, but here it can occurs
due to the possible existence of a static friction. \ Then, the
increments of the center of mass velocities and angular velocities
up to the value of the impulse $I_f$ at any moment $t_f$ within this
final \ pure rotation period  can be written in the form
\begin{eqnarray}
m_p\Delta \mathbf{v}_{cmp}^{(rp)}(I_f) &=&I_f\text{ }\mathbf{t}_3+\mathbf{I}%
^{fr},  \label{delvRP} \\
m_b\Delta \mathbf{v}_{cmp}^{(rp)}(I_f) &=&-I_f\text{ }\mathbf{t}_3-\mathbf{I}%
^{fr}, \\
I_p\Delta \mathbf{w}_p^{(rp)} &=&(\mathbf{r}_c-\mathbf{r}_p)\times \mathbf{I}%
^{fr}, \\
\widehat{\mathbf{I}}_b\cdot \Delta \mathbf{w}_b^{(rp)}(I_f) &=&-\mathbf{r}%
_c\times \mathbf{t}_3I_f-\mathbf{r}_c\times \mathbf{I}^{fr}, \\
\mathbf{I}_{imp}^{fr} &=&I_1^{fr}\mathbf{t}_1+I_2^{fr}\mathbf{t}_2.
\label{delwRP}
\end{eqnarray}

The parameters $I_f,I_1^{fr}$and $I_2^{fr}$ are the impulses of the
normal and frictional forces produced by the bat on the ball
starting from the time $t_{rp}$ up to the instant  $t_f$, and are
defined by the integrals
\begin{eqnarray}
I_f(t_f) &=&\int_{t_{rp}}^{t_f}N_{b\rightarrow p}(t)\text{ }dt, \\
I_1^{fr}(t_f) &=&\int_{t_{rp}}^{t_f}fr_{b\rightarrow p}(t)\text{ }dt, \\
I_2^{fr}(t_f) &=&\int_{t_{rp}}^{t_f}fr_{b\rightarrow p}(t)\text{ }dt.
\end{eqnarray}

A complete  set  equations for determining the evolution of all the
quantities  can now be defined, after assuming that we are
effectively in the $pure$ $rotation$ regime occurring before the
ending of the shock. \ The additional conditions for completely
fixing  the solution of the equations of motion  are basically the
vanishing of the two tangential components of the relative
velocities of  the contact points  at any moment during this ending
process. This condition for the relative velocity at the tangential
point can be written as follows
\begin{eqnarray}
0 &=&\Delta \mathbf{v}_{cmp}^{(rp)}(I_f)\cdot \mathbf{t}_i-\Delta \mathbf{v}%
_{cmb}^{(rp)}(I_f)\cdot \mathbf{t}_i+\Delta \mathbf{w}_p^{(rp)}(I_f)\mathbf{%
\times (\mathbf{r}_c-\mathbf{r}_{cmp})}\cdot \mathbf{t}_i- \\
&&\Delta \mathbf{w}_b^{(rp)}(I_f)\mathbf{\times \mathbf{r}_c}\cdot \mathbf{t}%
_i,\text{ \ }  \nonumber \\
i &=&1,2.  \nonumber
\end{eqnarray}

In these two equations, all the increments in the velocities can be
substituted in terms of the just defined  three values of the
impulses. They in turns can be solved for the two values of the
frictional impulses in terms of the \ unique values of the normal
force impulse as follows
\begin{eqnarray}
I_i^{fr}(t_f) &=&I_f\sum_{j=1,2}S_{ij}v_j,\text{ \ } \\
S_{ij} &=&\frac{\mathbf{t}_i\cdot \mathbf{t}_1\mathbf{t}_1\cdot \mathbf{t}_j%
}{D_1}+\frac{\mathbf{t}_i\cdot \mathbf{t}_2\mathbf{t}_2\cdot \mathbf{t}_j}{%
D_2}, \\
D_1 &=&\frac 1{m_b}+\frac 1{m_p}+\frac{(\mathbf{r}_c-\mathbf{r}_p)^2}{I_p}+%
\frac{(\mathbf{r}_c)^2}{I_t}+(\frac 1{I_3}-\frac 1{I_t})(\mathbf{k}\times
\mathbf{r}_c)^2, \\
D_2 &=&\frac 1{m_b}+\frac 1{m_p}+\frac{(\mathbf{r}_c-\mathbf{r}_p)^2}{I_p}+%
\frac{(\mathbf{r}_c)^2}{I_t}-\frac 1{I_t}(\mathbf{r}_c\cdot \mathbf{t}_2)^2.
\end{eqnarray}

Therefore, all the velocity increments in  (\ref{vpPR}-\ref{wbPR})
can be expressed as linear functions of the impulse done by the
normal force of the bat on the ball $I_f$. Note that this impulse is
defined as the one  transmitted after the instant in which the
system arrives to the pure rotation state up to any moment $t_f$. In
explicit form the values of the velocities as functions of $I_f$ are
defined by  equations (\ref{delvRP}-\ref{delwRP}) in the form
\begin{eqnarray}
\mathbf{v}_{cmp}^{(rp)}(I_f) &=&\mathbf{v}_{cmp}^{(rp)}(0)+\frac{I_f}{m_p}(%
\mathbf{t}_3+\sum_{i=1,2}\sum_{j=1,2}S_{ij}v_j\mathbf{t}_i),  \label{vpRPsol}
\\
\mathbf{v}_{cmp}^{(rp)}(I_f) &=&\mathbf{v}_{cmp}^{(rp)}(0)-\frac{I_f}{m_b}(%
\mathbf{t}_3+\sum_{i=1,2}\sum_{j=1,2}S_{ij}v_j\mathbf{t}_i),  \label{vbRPsol}
\\
\mathbf{w}_p^{(rp)}(I_f) &=&\mathbf{w}_p^{(rp)}(0)+\frac{I_f}{I_p}(\mathbf{r}%
_c-\mathbf{r}_p)\times (\sum_{i=1,2}\sum_{j=1,2}S_{ij}v_j\mathbf{t}_i),
\label{wpRPsol} \\
\mathbf{w}_b^{(rp)}(I_f) &=&\mathbf{w}_b^{(rp)}(0)-I_f\text{ }\widehat{%
\mathbf{I}}_b^{-1}\cdot (\mathbf{r}_c\times
\mathbf{t}_3+\mathbf{r}_c\times
\sum_{i=1,2}\sum_{j=1,2}S_{ij}v_j\mathbf{t}_i).  \label{wbRPsol}
\end{eqnarray}

In terms of these velocities, the total amount of kinetic energy
$E_f(I_f)$ at a fixed value of the impulse  $I_f$  is defined by the
formula
\begin{eqnarray}
E_f(I_f) &=&E^{(rp)}+\frac{m_p}2\left( 2\mathbf{v}_{cmp}(I_{rp})\cdot \Delta
\mathbf{v}_{cmp}^{(rp)}(I_f)+(\Delta \mathbf{v}_{cmp}^{(rp)}(I_f))^2\right) +
\nonumber \\
&&\frac{m_b}2\left( 2\mathbf{v}_{cmb}(I_{rp})\cdot \Delta \mathbf{v}%
_{cmb}^{(rp)}(I_f)+(\Delta \mathbf{v}_{cmb}^{(rp)}(I_f))^2\right) +
\nonumber \\
&&\frac{I_p}2\left( 2\mathbf{w}_p(I_{rp})\cdot \Delta \mathbf{w}%
_p^{(rp)}(I_f)+(\Delta \mathbf{w}_p^{(rp)}(I_f))^2\right) +  \nonumber \\
&&\frac 12{\Large (}2(\mathbf{w}_b(I_{rp})\cdot \widehat{\mathbf{I}}_b\cdot
\Delta \mathbf{w}_b^{(rp)}(I_f)+\Delta \mathbf{w}_b^{(rp)}(I_f)\cdot
\widehat{\mathbf{I}}_b\cdot \Delta \mathbf{w}_b^{(rp)}(I_f){\Large )}.
\label{Ef}
\end{eqnarray}

This completes the solution for the evolution equations for the
second process in which pure rotation occurs. \ In order to  fully
define the solution it only rests to determine the separation point.
However, it needs for a clear definition about the concrete
mechanisms of energy dissipation in the system. However, as it was
already remarked, one helpful outcome is that, under the assumed
rigidity assumptions,  no matter the form of the existing
dissipation  mechanisms, their differences only can alter a single
parameter of the problem: the particular value of the total impulse
done by the normal force of the bat on the ball at the just end of
the whole impact.

Let us discuss below the condition to be imposed for determining the
moment in which  the ball and the bat start to separate. It can be
constructed in terms of the energy balance in the system. The
separation point can be defined that one at which  the initial total
mechanical energy minus all the energy looses that had occurred  up
to this point,  just becomes equal to the  total kinetic energy  of
the ball and the bat. In explicit terms, it can be written in the
form
\begin{equation}
E^{(in)}+W_{fr}(I_{rp})+W_{add}(I_{rp}+I_{out})=E_f(I_{out})
\label{separation}
\end{equation}
where the total dissipative work of the friction $W_{fr}$ is given by the
general expression (\ref{wfriction}) after substituting $I=I_{rp}$%
\begin{equation}
W_{fr}(I_{rp})=-\mu \int_0^{I_{rp}}\mid \mathbf{v}\mid dI,
\label{trabfriction}
\end{equation}
and the losses due to other sources of dissipation in addition to
friction up the total value of the impulse done by the normal force
\ at the separation point $I_{rp}+I_{out},$ is the term
$W_{add}(I_{rp}+I_{out}).$ \ The right hand side of the equation is
the total kinetic energy of the ball and the bat at the separation
instant. That is, the kinetic energy at the point in which pure
rotation is established plus the increment due to the changes in all
the velocities during the  pure rotation evolution up to the
separation point. This kinetic energy is defined by equation
(\label{Ef}). \

Without specifying the nature of the additional sources of
dissipation in addition to the frictional one is not possible \ to
define the separation point.

Let us now consider by a moment  that the only energy dissipating
source is the friction. Then the additional losses term
$W_{add}(I_{rp}+I_f)$ vanishes. \ In this case relation
(\ref{separation}) becomes a simple quadratic equation for $I_{out}$
\ after substituting \ the expressions (\ref{vpRPsol}-\ref{wbRPsol})
for
the velocity increments, which all are homogeneous linear functions of $%
I_{out.}.$ In equation (\ref{separation}) all the quantities $E^{(in)},W_{fr}(0),$ $%
E^{(rp)},\mathbf{v}_{cmp}(I_{rp})$\textbf{, \ }$\mathbf{v}_{cmb}(t_{rp}),%
\mathbf{w}_p(I_{rp})$ and $\mathbf{w}_b(I_{rp})$ are already known
from the solutions just obtained in the previous stage of the shock.
\ Henceforth, the complete analytic solution of the shock problem
follows in the considered case in which $pure$ $rotation$ is
attained and friction is the only source of energy losses. The
situation of the existence of additional sources of losses will be
discussed in next section.

\subsection{A criterium for determining the shock case from the initial
conditions when dissipation is only frictional}

It seems useful to present a condition allowing to determine in
advance which kind of state will show the ball and the bat at the
end of the shock event, by only knowing the initial conditions.
Clearly, again, under the assumption of  existence of additional non
frictional sources of dissipation, such a criterium is expected to
depend on the specific mechanism to be considered. Therefore, we
will consider here only the case in which  the unique sources of
dissipation is determined by the friction. \

In this case the criterium is directly given by the sign of the quantity
\begin{equation}
C(\mathbf{v}_{cmp}^{(in)},\mathbf{v}_{cmb}^{(in)},\mathbf{w}_p^{(in)},%
\mathbf{w}_b^{(in)}\mathbf{)}=E^{(in)}(\mathbf{v}_{cmp}^{(in)},\mathbf{v}%
_{cmb}^{(in)},\mathbf{w}_p^{(in)},\mathbf{w}_b^{(in)}\mathbf{)}%
+W_{fr}(I_{rp})-E^{(rp)},  \label{decision }
\end{equation}
where $W(0)$ and $E^{(rp)}$ are implicit functions of the initial velocities
$\mathbf{v}_{cmp}^{(in)},\mathbf{v}_{cmb}^{(in)},\mathbf{w}_p^{(in)},\mathbf{%
w}_b^{(in)}$ , which were obtained in the course of the previous discussion.
\ If the sign of $C$ is positive, then the total energy after the bat and
the ball arrives to the $pure$ $rotation$ state $E^{(in)}(\mathbf{v}%
_{cmp}^{(in)},\mathbf{v}_{cmb}^{(in)},\mathbf{w}_p^{(in)},\mathbf{w}_b^{(in)}%
\mathbf{)}+W_{fr}(I_{rp})$ is larger than the kinetic energy of the two
bodies in this same state $E^{(rp)}.$ \ Therefore, at this moment the system
has energy stored in the form elastic deformations, and a non vanishing
normal force should remain existing. \ That is, the shock is not yet ended
and the second kind of the solution should be considered.

In another hand, if $C$ is negative, it indicates that the total
energy of the system after attaining $pure$ $rotation$ results to be
smaller than the kinetic energy in the same $pure$ $rotation$ state.
This means either, that energy is not being conserved in the
process, or that the system could not in fact attains the $pure$
$rotation$ state, and the shock ended with sliding contact points. \
In this case, the kind of solution to employ should be the one
discussed in section II. The case $C=0$ indicates that the shock
ends precisely at the moment in which the system arrives to the
$pure$ $rotation$ state.

\section{Description of experimental measures}

In this section we will  consider the application of the results
presented before to the description of the measures relative to the
scattering of a ball by a  bat in  Ref. \cite{ nathan}. The
experiment consisted in dropping  a free falling ball at certain
height which determines a vertical velocity of $ 4.0$ $\
\frac{\text{m}}{\text{sec}}$ at the instant  in which it shocks with
an horizontally oriented static bat. Then, a high  video recording
of the free fall of the ball allowed the authors to determine all
the kinematic  parameters before and after the impact. \ Experiments
were done, in which the ball was pitched a number of times with
fixed values of the angular velocities along the symmetry
axis of the bat. Experiences  for three values of the angular velocities  $%
w_o=79,$ $0,$ $-72$ $\frac{\text{rad}}{\sec }$ were done. The bat
has a barrel diameter of $6.67\ $cm,  a length  of $84$ cm and a
mass of $0.989$ kg. \ The center of mass of the bat is situated at
$26.5$ cm of  its barrel end. \ The ball for which measures were
done landed on the bat at distances along the bat axis ranging
between $14-16$ cm from the barrel end. Then, we will describe the
shocks by assuming that the  ball impacted the bat at an axial
distance of $15$ cm from  the barrel end.

The\ parameters of the  bat and the ball considered  in that work
determine the following \ values for the magnitudes defined in the
previous sections
\begin{eqnarray}
m_p &=&0.145\text{ \ kg ,}  \nonumber \\
m_b &=&0.989\text{ \ kg ,}  \nonumber \\
r_p &=&\sqrt{(\mathbf{\mathbf{r}_c-\mathbf{r}_{cmp})}^2}=0.036\text{
\  m}
\nonumber \\
r_b &=&0.03335\text{ m}  \nonumber \\
I_p &=&(2/5)(0.036)^2\text{ }m_p\text{ \ kg m}^{\text{2}},
\nonumber \\
I_t &=&0.0460\text{ \ kg m}^{\text{2}},  \nonumber \\
I_3 &=&4.39\times 10^{-4}\text{ \ kg m}^{\text{2}},  \nonumber \\
\mu  &=&0.5
\end{eqnarray}

The set of unit vectors sitting at the tangential point of the bodies were
chosen in the following way
\begin{eqnarray}
\ \mathbf{t}_1 &=&\cos (\theta )\text{ }\mathbf{j}-\sin (\theta )\text{ }%
\mathbf{i},  \nonumber \\
\mathbf{t}_2 &=&\mathbf{k},  \nonumber \\
\mathbf{t}_3 &=&\sin (\theta )\text{ }\mathbf{j}+\cos (\theta )\text{ }%
\mathbf{i},  \nonumber \\
\mathbf{r}_c &=&0.115\text{ }\mathbf{k}+0.03335\text{ }(\sin (\theta )\text{
}\mathbf{j}+\cos (\theta )\text{ }\mathbf{i}),
\end{eqnarray}
which reflect the fact that the barrel of the bat has been fixed as
a cylinder of  radius $0.03335$ cm and that the center of mass of
the bat has a minimum distance of $0.115$ m from the plane being
transversal to the bat axis and contains the contact point.   The
angle $\theta$ is the one formed between a radius traced from the
axis of the bat to the contact point. For the experimental
arrangement, the initial velocities of the ball and the bat just an
instant before the impact \ become
\begin{eqnarray}
\mathbf{v}_{cmp}^{(in)} &=&(-4,0,0),  \nonumber \\
\mathbf{v}_{cmb}^{(in)} &=&(0,0,0),  \nonumber \\
\mathbf{w}_p^{(in)} &=&(0,0,w_o),  \nonumber \\
\mathbf{w}_b^{(in)} &=&(0,0,0).
\end{eqnarray}

\subsection{ The solution of the shock problem for nearly vanishing impact
parameter }

In starting, let us exemplify the solution of the impact problem \
for the situation in which the vertically falling ball makes
contacts with the horizontally oriented bat at a very small value of
the impact parameter. That is, when the vertical line of falling
passes very close to the bat symmetry axis. \ For concreteness let
us suppose that the impact occurs at the small value of the angle
$\theta =\frac \pi {400}.$ This configuration will serve \ two
purposes  of the presentation. In first place it will illustrate \
the  application of the formal solutions found in  previous sections
\ to a concrete shock process. In second hand this particular
solution for scattering at zero impact parameter  case will serve
for phenomenologically constructing a description of the
experimental data \ presented in Ref. \cite{nathan}.   Firstly,
consider the solution of the equations (\ref{vtan}) for the
evolution for the tangential velocities in the firstly occurring
sliding period. \ The evaluation of the \ parameters $s_1,$ $s_2$
and $s_o$ defined in relations (\ref{s1}-\ref{s0}) leads to the
explicit form of the equations
\begin{eqnarray}
\frac{dv_1(I)}{dI} &=&-13.985044913377923\frac{v_1(I)}{\sqrt{%
v_1^2(I)+v_2^2(I)}}, \\
\frac{dv_2(I)}{dI} &=&-12.5866160651433\frac{v_2(I)}{\sqrt{v_1^2(I)+v_2^2(I)}%
}-0.08337500, \\
v_1(0) &=&0.0314156035548453\text{ \ }\frac{\text{meter}}{\sec }, \\
v_2(0) &=&0.
\end{eqnarray}

The employed initial values of the components of the tangent velocities $%
v_1(0)$ and $v_2(0)$ (along the vectors $ \mathbf{t}_1$ and \
$\mathbf{t}_2$ , respectively) were determined by projecting the
relative velocity at the just beginning of the impact, which is defined by equation (\ref{vtangente}%
), on each of these vectors. Note that the \ starting value of the tangent
velocity is small due to the assumptions \ of vanishing angular velocity of
the ball in common with the very small selection of the \ impact parameter.
\begin{figure}[h]
\hspace*{-0.4cm} \includegraphics[width=7.5cm]{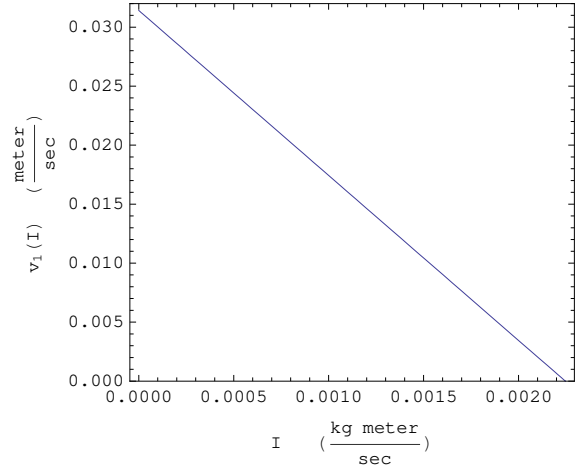}
 \caption{ The figure shows the evolution with
 the impulse $I$ of the component $v_1(I)$ of the tangent
 relative velocity  between the contact points on the bat and the
ball. It was evaluated by solving the corresponding differential
equations} \label{v1}
\end{figure}

\begin{figure}[h]
\hspace*{-0.4cm} \includegraphics[width=7.5cm]{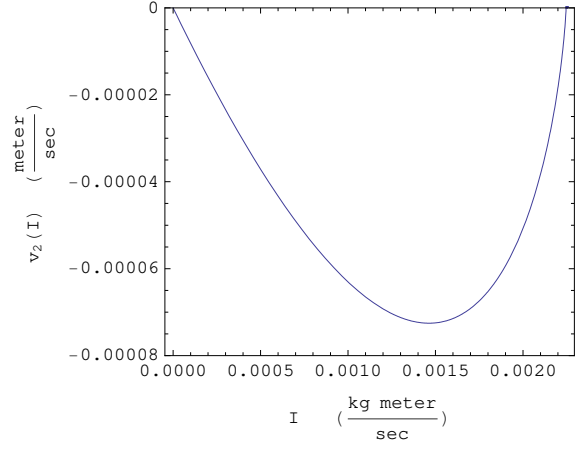}
 \caption{ The
figure shows the evolution with
 the impulse $I$ of the component $v_2(I)$ of the relative velocity  between
the contact points on the bat and the ball. It was evaluated by
solving the corresponding differential equations } \label{v2}
\end{figure}

\begin{figure}[h]
\hspace*{-0.4cm} \includegraphics[width=7.5cm]{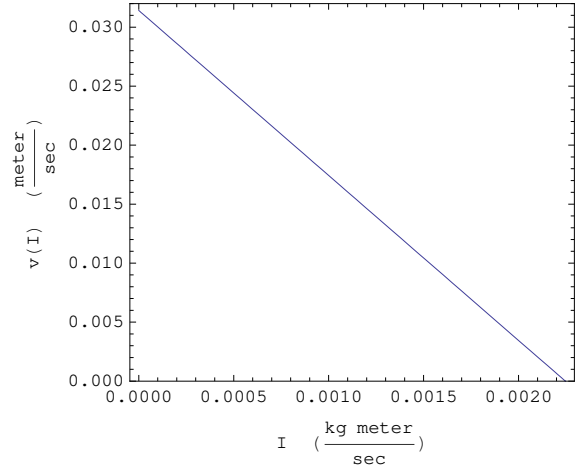}
 \caption{ The
figure shows the evolution with
 the impulse $I$ of the modulus  $v(I)$ of the tangent relative velocity
between the contact points in the bat and the ball. It was also
evaluated by solving the corresponding differential equations}
\label{v}
\end{figure}

 The solutions of these equations for $v_1(I)$, $v_2(I)$ and \
modulus of the tangent velocity $v(I)=\sqrt{v_1^2(I)+v_2^2(I)}$ \
are depicted in the figures \ref{v1}-\ref{v}. \ Note that the \
$v_2$ component is very  small, although non vanishing, which is
consistent with the fact that the shock is not strictly two
dimensional, because the  center of mass of the bat is out of the
plane which is orthogonal to the symmetry axis of the bat and passes
through the center of mass of the ball. \ The $v_2$ component,
although initially vanishing, develops values which grow up to a \
maximal one for tending to zero again. \ On another hand the \ $v_1$
component of velocity start decreasing from the start
to vanish exactly at the same value of the impulse $I$, for which the $%
v_2$ component also becomes equal to zero.   Therefore, the system
of equations predict that both components simultaneously tend to
approach a vanishing value. This property is exhibited by all the
solutions  of the scattering problem  found in this work to describe
the  experimental results in  Ref. (\cite{nathan}). \ The figure
\ref{v}  clearly illustrates the vanishing of the \ modulus of the
tangent velocity.  From the figures \ref{v1}-\ref{v} it can be seen
that the value of the impulse $I_{rp}$ of the normal force on the
ball $\ $ for which  the system  arrives to pure rotation for this
special scattering configuration  is
\begin{equation}
I_{rp=}0.002247237486554171\text{ \ \ }\frac{\text{kg meter}}{\sec }.
\label{Irp}
\end{equation}
\ \ \

Having found the evolution of the tangent velocities with the
variation of the impulse of the normal forces $I,$ we become able to
check whether the shock process will end in pure rotation state or
in a sliding condition between the contact points of the bat and the
ball. For this purpose let us evaluate the relation
(\ref{decision2}) by substituting the above defined
data for the velocities valid for the experiment and evaluating $%
W_{fr}(I_{rp})$  and  $E^{(rp)}$ through their respective formulae
(\ref {trabfriction}) and (\ref{Erp}). \ The evolution of the four
velocities of the ball and the bat from the starting of the shock up
to the moment in which the impulse at which $pure$ $rotation$ could
be attained, was evaluated from the formulae
(\ref{vpsol}-\ref{wbsol}) after  determining the  impulse of the
friction  from its definition (\ref{frictionImpulse}).

The evaluation results in a  positive value for the $C$  function
\begin{eqnarray}
C(\mathbf{v}_{cmp}^{(in)},\mathbf{v}_{cmb}^{(in)},\mathbf{w}_p^{(in)},%
\mathbf{w}_b^{(in)}\mathbf{)} &=&E^{(in)}(\mathbf{v}_{cmp}^{(in)},\mathbf{v}%
_{cmb}^{(in)},\mathbf{w}_p^{(in)},\mathbf{w}_b^{(in)}\mathbf{)}%
+W_{fr}(I_{rp})-E^{(rp)}  \nonumber \\
&=&0.00896798\text{ \ \ Joules,} \\
E^{(in)}(\mathbf{v}_{cmp}^{(in)},\mathbf{v}_{cmb}^{(in)},\mathbf{w}_p^{(in)},%
\mathbf{w}_b^{(in)}\mathbf{)} &=&1.16 , \\
E^{(rp)} &=&1.15101\text{ \ \ Joules} , \\
W_{fr}(I_{rp}) &=&-0.0000176437\text{ \ \ Joules\ }.
\end{eqnarray}

Therefore, assumed that the only source of energy losses is the
friction dissipation, since $C$ results to be positive at the
instant in which the transmitted  impulse is $I_{rp}$ for which
$pure$ $rotation$ is attained, the total mechanical energy is larger
that the kinetic energy $E^{(rp)}$ of the two bodies. \  In addition
the values of the dissipated energy by the friction
$W_{fr}(I_{rp})$\ results to be very small when $pure$ $rotation$ is
established. Since the kinetic energy $E^{(rp)}$ at pure rotation \
has a value close to the total initial mechanical energy, it is
clear that the pure rotation state is attained just at the very
beginning of the impact. This is compatible with the very small
impulse \ $I_{rp}$ (see \ref{Irp}) transmitted by the normal force
in arriving to vanishing sliding of the contact surfaces.

Having defined that the scattering situation corresponds to a pure
rotation final state, let us consider  now the  evolution of the
system after the impulse done by the normal force continues to be
growing  under the pure rotation state. The change of all the
velocities of the bat and the ball in this new \ process \ are
determined as simple linear functions of  the impulse $I_f$\ by
equations   (\ref {vpRPsol}-\ref{wbRPsol}). \

Let us consider first the case in which only friction is able to
dissipate energy. Then the condition for the separation of the ball
and the bat  (\ref {separation}) gives a simple quadratic equation
for the determination of the value of  $I_f$ at which the two bodies
separate. The condition for separation and its explicit  form are
\begin{eqnarray}
E^{(in)}+W_{fr}(I_{rp}) &=&E_f(I_f) \\
1.1599823562526097 &=&1.1510143769488514+  \nonumber \\
&&I_f\text{ }(-3.9814607487573555`+4.09744896389227\text{ }I_f),
\label{Ifsol}
\end{eqnarray}
which  give for the value of the impulse of the normal force during
the pure rotation interval up to the separation point, assumed that
the only existing energy  losses are associated to friction, the
result
\begin{equation}
I_f^{(out,1)}=0.97394\text{ \ }\frac{\text{kg meter}}{\sec }.
\end{equation}

This value compared  with $I_{rp=}0.00224723748$ \ evidences that
the pure rotation state  was directly established at the beginning
of the impact.
\begin{figure}[h]
\hspace*{-0.4cm} \includegraphics[width=7.5cm]{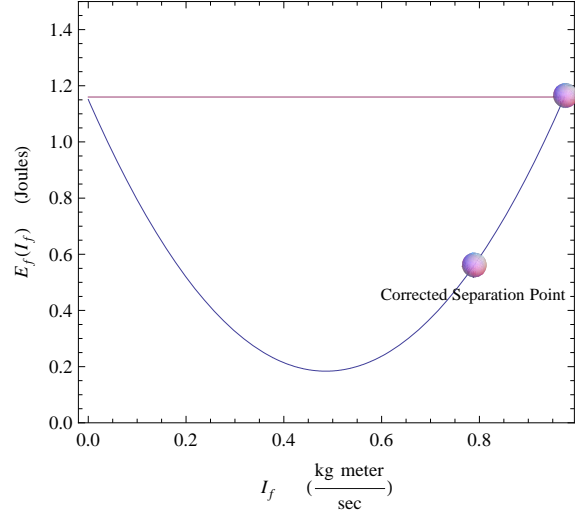}
\caption{ The
plot of the total kinetic energy of the bat and the ball $E_f$ as a
function of the  impulse of the normal force of the bat on the ball
$I$ in the ending pure rotation process. The horizontal line defines
the conserved value of the total mechanical energy after the pure
rotation state is established. The difference between the horizontal
line and $E_f(i)$ gives the amount of energy stored in the form of
elastic deformation at any  value of the impulse $I$. The ball
depicted on the horizontal indicates  the separation point between
the ball and the bat when the only source of dissipation is the
friction. The similar ball laying on the curve of $E_f$ indicates
the separation point to be defined  by the model here constructed
for evaluating the non frictional energy losses. } \label{Ef}
\end{figure}

The determination of $I_f$ \ finishes the solution of the shock
problem in this case, since substituting this value in  the
expressions (\ref {vpRPsol}-\ref{wbRPsol}) \ determines all the
center of mass  and  angular velocities of the bat and the ball at
the separation instant.

\subsubsection{ Consideration of the losses due to inelastic processes}

Let us consider now the situation in which there exist energy
dissipation sources in addition to the friction. For this purposes
consider figure  \ref{Ef}  which illustrates the solution the just
discussed solution. \ The parabolic curve  shows the \ dependence of
the  kinetic energy $E_f$ as a function of $I_f.$  The horizontal
line indicates the value of the conserved  mechanical energy
$E^{(in)}+W_{fr}(I_{rp})$ in the considered pure rotation period. \
The figure shows how, as the system evolves  from the instant in
which $pure$ $rotation$ was established, it  accumulates energy in
elastic form as signaled by the difference between the kinetic
energy $E_f$ and the conserved  mechanical energy
$E^{(in)}+W_{fr}(I_{rp}),$ up to a maximum value, which afterwards
starts to decrease. \ This behavior will be taken into account in
what follows to construct a model for the  non elastic dissipation
processes different from frictional one. The basic purpose will be
to apply the analysis to the description of the measures of
scattering of  a ball by a bat  given in Ref. \cite{nathan}\ .

Assume that we are already in the pure rotation state, as the former
evaluations in this section had stated. Then, let us consider the
more general condition \ (\ref{separation} ) for determining the
separation  point between the ball and bat, in which the additional
losses term $W_{add}(I_{rp}+I_f)$ was introduced. \  It is clear
that the function $W_{add}(I_{rp}+I_f)$ depends on the types of
materials constituting the bat \ and the ball, in particular on
their properties under the large local deformations occurring \ near
the impact point. \ Therefore, we have not at hand well defined
information about how the non elastic dissipation is occurring as
the impulse of the normal force is growing when the shock develops.
\ Therefore, we will  employ a global condition for the
determination of \ the amount of dissipation in addition to the
frictional ones. \ As remarked before, this condition was suggested
by the data depicted in figure \ref{Ef}.

The condition adopted an  intuitively motivated notion:  that the amount of
 non elastic losses in any type of  shock will be given by a fixed fraction
$e^2$ of the maximal amount of elastic energy which is accumulated
along the evolution of the system, when  dissipation is only given
by friction. In explicit terms
\begin{equation}
W_{add}(I_{rp}+\ I_f^{(out,2)})=-e^2(E^{(in)}+W_{fr}(I_{rp})-E_f(I_f^{\max
})),
\end{equation}

That is, the additional energy losses at the value of the impulse at
which the bodies separate $I_f^{(out,2)},$ will be chosen to be a
fraction $e^2$ of the  difference between the total mechanical
energy after pure rotation is attained $E^{(in)}+W_{fr}(I_{rp})$ (a
quantity which is conserved in the assumed case in the above
definition of pure frictional losses) and the
total kinetic energy $E_f(I_f^{\max })$ at the value of the impulse $%
I_f^{\max }$ . This value $I_f^{\max }$ correspond to the impulse at which
the stored elastic energy $(E^{(in)}+W_{fr}(I_{rp})-E_f(I_f))$ is maximal as
a function of $I_f$ when pure frictional dissipation is assumed. Then the
condition for separation (\ref{separation}) \ gets the general form
\begin{equation}
E_f(I_f^{(out,2)})=E^{(in)}+W_{fr}(I_{rp})-e^2(E^{(in)}+W_{fr}(I_{rp})-E_f(I_f^{\max })),
\label{condition2}
\end{equation}
from which the value of $I_f^{(out,2)}$ \ can be directly obtained because $%
E_f(I_f^{(out,2)})$ \ is a quadratic function of $I_f^{(out,2)}$
defined by (\ref{Ifsol}).

Once the value of $I_f^{(out,2)}$ is at hand,  its substitution  in
(\ref {vpRPsol}) and (\ref{wpRPsol}) allows to evaluate for  the
absolute values of the ball center of mass and angular velocities
which are basic quantities measured in \ Ref. (\cite{nathan} ),  the
expressions
\begin{eqnarray}
|\mathbf{v}_p^{fin}| &=&\sqrt{\mathbf{v}_{cmp}^{(rp)}(I_f^{(out,2)})\cdot
\mathbf{v}_{cmp}^{(rp)}(I_f^{(out,2)})},  \label{finalb} \\
\mathbf{v}_{cmp}^{(rp)}(I_f^{(out,2)}) &=&\mathbf{v}_{cmp}^{(rp)}(0)+\frac{%
I_f^{(out,2)}}{m_p}(\mathbf{t}_3+\sum_{i=1,2}\sum_{j=1,2}S_{ij}v_j\mathbf{t}%
_i), \\
|\mathbf{w}_p^{fin}| &=&\sqrt{\mathbf{w}_p^{(rp)}(I_f^{(out,2)})\cdot
\mathbf{w}_p^{(rp)}(I_f^{(out,2)})}, \\
\mathbf{w}_p^{(rp)}(I_f^{(out,2)}) &=&\mathbf{w}_p^{(rp)}(0)+\frac{%
I_f^{(out,2)}}{I_p}(\mathbf{r}_c-\mathbf{r}_p)\times
(\sum_{i=1,2}\sum_{j=1,2}S_{ij}v_j\mathbf{t}_i).
\end{eqnarray}

These quantities were calculated  for a set of values of the impact
parameter defined as the minimal distance between vertical line
along which the center of the ball was falling and the initially
horizontally oriented symmetry axis of the bat.

\ The value of the constant $e^2$ was determined by fixing the
measured value of the output center of mass velocity in Ref. \cite
{nathan}, at the particular condition of scattering considered at
the beginning of this section. That is when the angular velocity of
the falling ball is zero and the center of mass velocity is 4.0
$\frac{meter}{\sec }$ at a nearly vanishing value of the impact
parameter. \  The condition (\ref {condition2}) \ for determining
the separation in this case gets the form \
\begin{eqnarray}
r_1+r_2\text{ }e^2 &=&r_3+I_f^{(out,2)}\text{ }(r_4+r_5\text{ }I_f^{(out,2)})
\\
r_1 &=&1.1599823562526097, \\
r_2 &=&0.976157, \\
r_3 &=&1.1510143769488514 \\
r_4 &=&-3.9814607487573555` \\
r_5 &=&4.09744896389227
\end{eqnarray}
where the values of the impulse at the minimum value of the kinetic energy $%
E_f$\ for the assumed set of initial data is $I_f^{\max }=0.485846$ and the
corresponding value of the kinetic energy at this point is $E_f(I_f^{\max
})=0.183825.$ \

The fixation of $e^2$ proceeded by assuming some trial values of
this quantity and solving the equation for $I_f^{(out,2)}$ for each
one of them, by further evaluating the absolute value of the  final
ball be velocity by using (\ref{finalb}). The trials were repeated
after to arrive to a final output velocity of the ball being around a value of $|\mathbf{v}%
_p^{fin}|=1.44629\ \frac{meter}{\sec },$ which is close to the \ one
measured in  Ref. \cite{nathan} for the assumed scattering
conditions.  The resulting \ value of $e^2$ was  $0.6183822.$
\begin{figure}[h]
\hspace*{-0.4cm} \includegraphics[width=7.5cm]{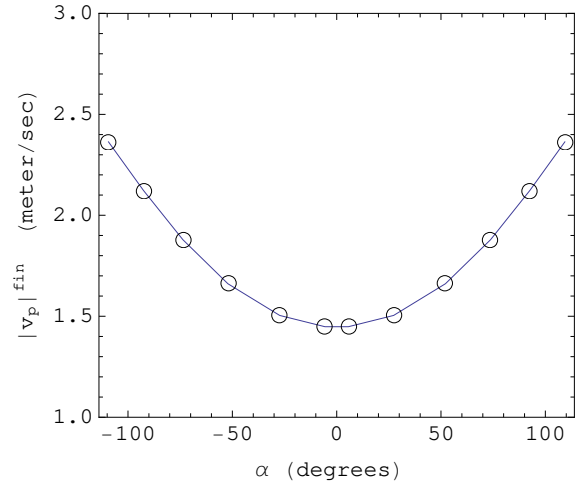} \caption{
The figure show the calculated  absolute value of the center of mass
velocity of the  ball after the impact as a function of the
scattering angle $\alpha$.  The value at $\alpha=0$ was fixed to be
close to the measured one in Ref. \cite{nathan}, which determined
the value of the constant $e^2$ defining the non-frictional energy
losses } \label{vpfin0}
\end{figure}

\  Once this parameter was determined, we performed evaluations \ of
the output angular and center of mass velocities of the ball for
various values of the angle $\theta.$  Different sets of evaluations
were done for these quantities, one for each of three values of the
initial angular velocity of the ball, for which measures
 were done in the experiments: \ $%
w_o=+79,0,-72$ \ $\frac{rad}{\sec }.$ The results for the  absolute
values of the  center mass velocities of the ball
$|\mathbf{v}_p^{fin}|$ were plotted as functions of the scattering
angle $\alpha $  (expressed in degrees) formed by the output ball
velocity and its corresponding input  value. This angle is defined
by
\begin{equation}
\alpha =\frac{180}\pi  \text{ArcSin}(-\frac{\mathbf{k}\cdot \mathbf{v}%
_p^{fin}\times \mathbf{i}}{|\mathbf{k}\cdot \mathbf{v}_p^{fin}\times \mathbf{%
i|}}).
\end{equation}

\begin{figure}[h]
\hspace*{-0.4cm} \includegraphics[width=7.5cm]{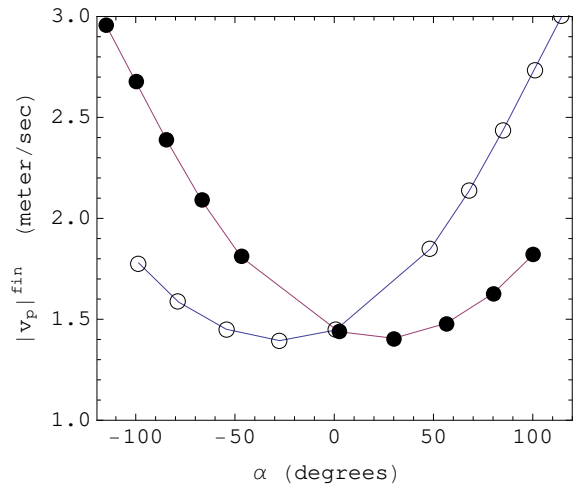}
\caption{ The figure shows two sets of calculated absolute values of
the center of mass velocity of the  ball after the impact as
functions of the scattering angle $\alpha$. The open circles depict
the velocity values when the initial angular velocity of the ball is
$ +79 \frac {\text{rad}}{sec}$.  The filled ones  indicate the
velocities for an initial angular velocity of $-72 \frac
{\text{rad}}{sec}$ } \label{vpfin7972}
\end{figure}
We also evaluated the ending angular velocity of the ball as
functions in this case of the impact parameter $E$ (expressed in
inches) defined by
\begin{equation}
E=(r_p+r_b)\cos (\theta).
\end{equation}

The results for the final absolute value of the ball velocity $|\mathbf{v}%
_p^{fin}|$  as a function of $\alpha $ when its initial angular
velocity is  taken as vanishing   are depicted in figure
(\ref{vpfin0}). \ As described before the  value of the final center
of mass velocity of the ball was  phenomenologically fixed  to
approximately  reproduce the measured value near $1.5$
$\frac{\text{meter}}{\sec }$ at $\alpha =0$ for the zero initial
angular momentum of the ball experiment. \ No other  parameter
fixation was  additionally done. \ Therefore all the shown data for
the values of the ball velocities in dependence the scattering angle
$\alpha $ represent  predictions of the  analysis done here. \ The
comparison of the  results  with the ones plotted in the
corresponding figure  (2) (top) of  Ref. \cite{nathan}  permits to
conclude  that the model solution found here satisfactorily
reproduces the  measured data. \
\begin{figure}[h]
\hspace*{-0.4cm} \includegraphics[width=8.5cm]{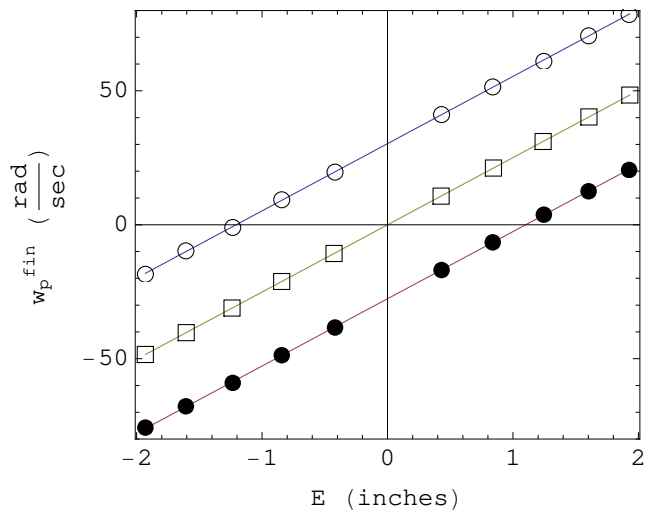}
\caption{ The figure illustrate the variation with the impact
parameter of three sets of values of the calculated angular velocity
of the ball after the shock. The open circles correspond to an
initial angular velocity equals to $w_o=+79 \frac {rad}{sec}$ . The
squares show the evaluated angular velocities for $w_o=0$.   The
filled circles indicate the calculated angular velocities for
$w_o=-72 \frac {rad}{sec}$  } \label{wpfin79072}
\end{figure}
\  The predicted values for the final center of mass velocities
of the ball for  the cases in which it shocks with a  4.0 $%
\frac{\text{meter}}{\sec }$ center of mass velocity with the static
and horizontal bat, and  having angular velocities of values $+79$ and $-72$\ $%
\frac{\text{rad}}{\sec }$, are presented in figure \ref{vpfin7972}.
These results again satisfactorily match  the corresponding
measurements shown in figure  3 (top) in  Ref.  \cite{nathan}. It
can be noted that the same, natural to be expected,  asymmetry of
the velocities with respect to the change of the sign of the
scattering angle $\alpha $  is exhibited and the quantitative values
also approach the measured  ones within the experimental errors.
Finally,  the ending  angular velocities  of the ball  for each of
the three values of the initial angular  velocities are plotted in
figure  \ref{wpfin79072}. \ In this case the nearly linear
dependences for the three experiments  measured in Ref.
\cite{nathan} and shown in figure 5 of that work, are satisfactorily
reproduced in slope and  values within the precision allowed by the
degree of dispersion of the measured values.

\section{ Summary}

We have presented a full solution  of the general problem of the
scattering between  spherical object and a cylindrically symmetric
one when both of them are assumed as perfectly rigid bodies and the
friction is assumed to show the  standard properties and being  the
only source of energy dissipation. \ A simple criterium is
determined allowing to decide  from the beginning whether the final
states of the bodies will correspond or not to  sliding contact
surfaces or to the contact points being at rest at the end. The \
exact solution for the evolution of all the physical quantities
during the shock is also found, when  other types of energy
dissipation in addition to the frictional one are present. \
However, in this case the only lacking information  is the concrete
determination of the value of net impulse done by the bat on the
ball at the separation point. The determination of this point needs
of detailed information on the  additional sources of dissipation.

A condition for determining this point is constructed in this work
in order to apply the results to the description of published
experimental measurements of the scattering of a ball by a bat. The
construction became suggested by the  possibility of properly
identifying the amount of elastic energy in the process of the
solution in the case of the existence of  only frictional losses.
The analysis is applied to determine the solution of the problem of
the scattering of a ball by a bat for which experimental measures
were presented in Ref. \cite{nathan}. \ The experience corresponded
to a vertically falling ball which impacts at 4.0
$\frac{\text{meter}}{\sec }$ an horizontally laying and static and
non rotating bat. \ The solution of the problem \ satisfactorily
reproduced the  measured  dependence of the final velocity of the
ball  as  function of the scattering angle. This happens for each of
the three values of the initial angular velocity of the  ball
employed  in the experiments. The behavior of the final angular
velocity of the ball on the  impact parameter for each of the cited
values of the initial angular velocity of the ball are also
appropriately described.

\begin{acknowledgments}
The authors acknowledge support received from  the Caribbean Network
on Quantum Mechanics, Particles and Fields (Net-35) of the ICTP
Office of External Activities (OEA), the "Proyecto Nacional de
Ciencias B\'{a}sicas" (PNCB) of CITMA, Cuba.  The helpful remarks on
the theme received from R. Homs, L. A. Delpino, A. del Pozo, J.
Gulin, E. Altshuler, Y. Nunez and M. Bizet are also greatly
acknowledged.

\end{acknowledgments}

\section{The rigid and frictionless shock solution}

In this appendix, we will consider the solution of the shock problem for the
case in which the interaction force between the bodies $\mathbf{F}(t)$
during the impact is conservative and normal. Due to its impact nature, let
us consider the force as given by a Dirac delta distribution
\[
\mathbf{F}(t)=\mathbf{I}_{imp}\mathbf{\delta }(t-t_0),
\]
in which $\mathbf{I}_{imp}$ is the total impulse vector transmitted by the
force. \ Then, the Newton equations for the problem can be written as
follows
\begin{eqnarray}
m_p\frac d{dt}\mathbf{v}_{cmp}(t) &=&\mathbf{I}_{imp}\mathbf{\delta }(t-t_0),%
\text{ \ \ \ \ \ }  \nonumber \\
\text{\ }m_b\frac d{dt}\mathbf{v}_{cmb}(t) &=&-\mathbf{I}_{imp}\mathbf{%
\delta }(t-t_0),  \nonumber \\
\widehat{\mathbf{I}}_p\cdot \frac d{dt}\mathbf{w}_p(t) &=&(\mathbf{r}_c-%
\mathbf{r}_{cmp})\times \mathbf{I}_{imp}\mathbf{\delta }(t-t_0),\text{ \ \ \
\ }  \nonumber \\
\text{\ }\widehat{\mathbf{I}}_b\cdot \text{ }\frac d{dt}\mathbf{w}_b(t) &=&-%
\mathbf{r}_c\times \mathbf{I}_{imp}\mathbf{\delta }(t-t_0).  \label{consys}
\end{eqnarray}

Note that the third equation was expressed in terms of the angular impulse
respect to the center of mass of the ball. This was done by using the
definition (\ref{angdef}) of the angular momentum of the ball respect to the
reference frame sitting at the center of mass of the bat, and the first of
the equations in (\ref{consys}). \ Integrating the above equations over
time, it follows
\begin{eqnarray}
m_p(\mathbf{v}_{cmp}^{(out)}-\mathbf{v}_{cmp}^{(in)}) &=&\mathbf{I}_{imp},%
\text{ \ \ \ \ \ }  \nonumber \\
\text{\ }m_b(\mathbf{v}_{cmb}^{(out)}-\mathbf{v}_{cmb}^{(in)}) &=&-\mathbf{I}%
_{imp},  \nonumber \\
\widehat{\mathbf{I}}_p\cdot \text{ (}\mathbf{w}_p^{(out)}-\mathbf{w}%
_p^{(in)}) &=&(\mathbf{r}_c-\mathbf{r}_{cmp})\times \mathbf{I}_{imp},\text{
\ \ \ \ }  \nonumber \\
\widehat{\mathbf{I}}_b\cdot \text{ (}\mathbf{w}_b^{(out)}-\mathbf{w}%
_b^{(in)}) &=&-\mathbf{r}_c\times \mathbf{I}_{imp},
\end{eqnarray}
where the superindices $(in)$ and $(out),$ indicate the values of the
magnitudes at an instant before and after the start of the shock,
respectively.

\ Let us consider now the condition satisfied by the impulse of the
interaction force in order to implement our two suppositions: conservation
of energy and the absence of friction between the contact surfaces. \ Its is
clear that if there is no friction between the contact planes there will be
no projection of the forces in the tangent planes and therefore:
\begin{eqnarray}
\mathbf{t}_i\cdot \mathbf{I}_{imp} &=&0=m_p(\mathbf{v}_{cmp}^{(out)}-\mathbf{%
v}_{cmp}^{(in)})\cdot \mathbf{t}_i,\text{ \ \ }i=1,2,  \nonumber \\
\mathbf{t}_i\cdot \mathbf{I}_{imp} &=&0=m_b(\mathbf{v}_{cmb}^{(out)}-\mathbf{%
v}_{cmb}^{(in)})\cdot \mathbf{t}_i,\text{ \ \ \ }i=1,2.
\end{eqnarray}

Thus, the tangent components of the center of mass velocities after the
shock are exactly the same as themselves before the impact. Therefore, these
two quantities  are already determined. \ For the normal to the tangent
plane of the center of mass components it follows,
\begin{equation}
m_p(\mathbf{v}_{cmp}^{(out)}-\mathbf{v}_{cmp}^{(in)})\cdot \mathbf{t}_3=-m_b(%
\mathbf{v}_{cmb}^{(out)}-\mathbf{v}_{cmb}^{(in)})\cdot \mathbf{t}_3,
\end{equation}
which coincides in form with the usual result for the simple collinear and
conservative shock between two bodies. \

For the ball, the simplification is stronger, because the impact force, as
having no tangent component, has a vanishing angular impulse
\begin{equation}
I_p\text{ (}\mathbf{w}_p^{(out)}-\mathbf{w}_p^{(out)})=-(\mathbf{r}_c-%
\mathbf{r}_{cmp})\times \mathbf{I}_{imp},=0,
\end{equation}
which directly implies that the angular velocity vector of the ball is
conserved during the shock:
\begin{equation}
\mathbf{w}_p^{(out)}=\mathbf{w}_p^{(in)},
\end{equation}
furnishing the solution for these variables after the shock is finished.

\ Further, the normal direction of the conservative impulsive force implies
that its angular impulse on the bat is directed in the $\mathbf{t}_1$
direction. This property, then implies the conservation of the components of
the initial angular velocity along the $\mathbf{t}_2$ $\ $and $\ \ \mathbf{t}%
_3$ spacial directions:
\begin{eqnarray}
\text{(}\mathbf{w}_b^{(out)}-\mathbf{w}_b^{(in)})\cdot \mathbf{t}_2 &=&0, \\
\text{(}\mathbf{w}_b^{(out)}-\mathbf{w}_b^{(in)})\cdot \mathbf{t}_3 &=&0.
\end{eqnarray}

The remaining two integrated Newton equations, constitute a set of two
equations for the yet undetermined variables $(\mathbf{v}_{cmb}^{(out)}-%
\mathbf{v}_{cmb}^{(in)})\cdot \mathbf{t}_3,$ ($\mathbf{w}_b^{(out)}-\mathbf{w%
}_b^{(in)})\cdot \mathbf{t}_1$\ and $(\mathbf{v}_{cmp}^{(out)}-\mathbf{v}%
_{cmp}^{(in)})\cdot \mathbf{t}_3,$ that can be written in the forms
\begin{eqnarray}
-m_b(\mathbf{v}_{cmb}^{(out)}-\mathbf{v}_{cmb}^{(in)})\cdot \mathbf{t}_3
&=&m_p(\mathbf{v}_{cmp}^{(out)}-\mathbf{v}_{cmp}^{(in)})\cdot \mathbf{t}_3,
\nonumber \\
\text{(}\mathbf{w}_b^{(out)}-\mathbf{w}_b^{(in)})\cdot \mathbf{t}_1 &=&-m_p(%
\mathbf{t}_1\times \mathbf{t}_2\cdot \mathbf{t}_3)(\mathbf{r}_{cmp}\cdot
\mathbf{t}_2)(\mathbf{v}_{cmp}^{(out)}-\mathbf{v}_{cmp}^{(in)})\cdot \mathbf{%
t}_3.  \nonumber
\end{eqnarray}

These equations state that the discontinuities in the normal and angular
velocities of the bat, are both expressed in terms of the discontinuity of
the normal velocity of the ball. Thus, after finding another equation being
able in determining this unique ball velocity discontinuity, the problem
will become solved.

This additional condition, should correspond to impose the conservation of
the energy after the end of the shock. \ Its expression is
\begin{eqnarray}
&&\frac{m_p}2\sum_{i=1}^3(\mathbf{v}_{cmp}^{(out)}\cdot \mathbf{t}_i)^2+%
\frac{I_p}2\sum_{i=1}^3(\mathbf{w}_p^{(out)}\cdot \mathbf{t}_i)^2+  \nonumber
\\
&&\frac{m_b}2\sum_{i=1}^3(\mathbf{v}_{cmb}^{(out)}\cdot \mathbf{t}%
_i)^2+\frac 12\sum_{i=1}^3I_i(\mathbf{w}_b^{(out)}\cdot \mathbf{t}_i)^2
\nonumber \\
&=&\frac{m_p}2\sum_{i=1}^3(\mathbf{v}_{cmp}^{(in)}\cdot \mathbf{t}_i)^2+%
\frac{I_p}2\sum_{i=1}^3(\mathbf{w}_p^{(in)}\cdot \mathbf{t}_i)^2+  \nonumber
\\
&&\frac{m_b}2\sum_{i=1}^3(\mathbf{v}_{cmb}^{(in)}\cdot \mathbf{t}_i)^2+\frac
12\sum_{i=1}^3I_i(\mathbf{w}_b^{(in)}\cdot \mathbf{t}_i)^2.  \label{conserv}
\end{eqnarray}

After using the known information about the variables  which have
been already determined, all the quantities entering these relation
can be expressed as functions of the only three remaining unknown
quantities in the following way:
\begin{eqnarray}
\mathbf{v}_{cmb}^{(out)} &=&\sum_{i=1}^3(\mathbf{v}_{cmb}^{(in)}\cdot
\mathbf{t}_i)\text{ \ }\mathbf{t}_i+((\mathbf{v}_{cmb}^{(out)}-\mathbf{v}%
_{cmb}^{(in)})\cdot \mathbf{t}_3)\text{ }\mathbf{t}_3  \nonumber \\
&=&\mathbf{v}_{cmb}^{(in)}+((\mathbf{v}_{cmb}^{(out)}-\mathbf{v}%
_{cmb}^{(in)})\cdot \mathbf{t}_3)\text{ }\mathbf{t}_3,  \nonumber \\
\mathbf{v}_{cmp}^{(out)} &=&\sum_{i=1}^3\mathbf{v}_{cmp}^{(in)}\cdot \mathbf{%
t}_i)\text{ \ }\mathbf{t}_i+((\mathbf{v}_{cmp}^{(out)}-\mathbf{v}%
_{cmp}^{(in)})\cdot \mathbf{t}_3)\text{ }\mathbf{t}_3  \nonumber \\
&=&\mathbf{v}_{cmp}^{(in)}+((\mathbf{v}_{cmp}^{(out)}-\mathbf{v}%
_{cmp}^{(in)})\cdot \mathbf{t}_3)\text{ }\mathbf{t}_3,  \nonumber \\
\mathbf{w}_p^{(out)} &=&\mathbf{w}_p^{(in)},  \nonumber \\
\mathbf{w}_b^{(out)} &=&\sum_{i=1}^3(\mathbf{w}_b^{(in)}\cdot \mathbf{t}_i)%
\text{ \ }\mathbf{t}_i+((\mathbf{w}_b^{(out)}-\mathbf{w}_b^{(in)})\cdot
\mathbf{t}_1)\text{ }\mathbf{t}_1  \nonumber \\
&=&\mathbf{w}_b^{(in)}+((\mathbf{w}_b^{(out)}-\mathbf{w}_b^{(in)})\cdot
\mathbf{t}_1)\text{ }\mathbf{t}_1.
\end{eqnarray}

Henceforth, the equations for the three remaining variables to be
determined, take the forms
\begin{eqnarray}
m_p(\mathbf{v}_{cmp}^{(out)}-\mathbf{v}_{cmp}^{(in)})\cdot \mathbf{n}_c
&=&-m_b(\mathbf{v}_{cmb}^{(out)}-\mathbf{v}_{cmb}^{(in)})\cdot \mathbf{n}_c,
\nonumber \\
I\text{(}\mathbf{w}_b^{(out)}-\mathbf{w}_b^{(in)})\cdot \mathbf{t}_1 &=&-m_p(%
\mathbf{r}_{cmp}\cdot \mathbf{t}_2)\text{(}\mathbf{v}_{cmp}^{(out)}-\mathbf{v%
}_{cmp}^{(in)})\cdot \mathbf{t}_3,  \nonumber \\
0 &=&\frac{m_p}2((\mathbf{v}_{cmp}^{(out)}\cdot \mathbf{t}_3)^2-(\mathbf{v}%
_{cmp}^{(in)}\cdot \mathbf{t}_3)^2)+\frac{m_b}2((\mathbf{v}%
_{cmb}^{(out)}\cdot \mathbf{t}_3)^2-(\mathbf{v}_{cmb}^{(in)}\cdot \mathbf{t}%
_3)^2)+  \nonumber \\
&&\frac 12I((\mathbf{w}_b^{(out)}\cdot \mathbf{t}_1)^2-(\mathbf{w}%
_b^{(out)}\cdot \mathbf{t}_1)^2).
\end{eqnarray}

After defining the three quantities
\begin{eqnarray}
x &=&(\mathbf{v}_{cmp}^{(out)}-\mathbf{v}_{cmp}^{(in)})\cdot \mathbf{t}_3,
\nonumber \\
y &=&(\mathbf{v}_{cmb}^{(out)}-\mathbf{v}_{cmb}^{(in)})\cdot \mathbf{t}_3,
\nonumber \\
z &=&(\mathbf{w}_b^{(out)}-\mathbf{w}_b^{(in)})\cdot \mathbf{t}_1,
\end{eqnarray}
the equations get the simple forms
\begin{eqnarray}
x &=&-\frac{m_b}{m_p}y,  \nonumber \\
z &=&-\frac{m_p}I(\mathbf{r}_{cmp}\cdot \mathbf{t}_2)x,  \nonumber \\
0 &=&\frac{m_p}2(2(\mathbf{v}_{cmp}^{(in)}\cdot \mathbf{t}_3)\text{ }x+x^2)+%
\frac{m_b}2(2(\mathbf{v}_{cmb}^{(in)}\cdot \mathbf{t}_3)\text{ }y+y^2)+
\nonumber \\
&&\frac 12I(2(\mathbf{w}_b^{(in)}\cdot \mathbf{t}_1)\text{ }z+z^2),
\end{eqnarray}
which after eliminating $y$ and $z$ give the following quadratic equation
for $x$
\begin{eqnarray}
&&x\left\{ \left( \frac{m_p}2+\frac{m_p^3}{2m_b^2}+\frac I2(\frac{m_p}I)^2(%
\mathbf{r}_{cmp}\cdot \mathbf{t}_2)^2\right) x\right. +  \nonumber \\
&&\left. m_p\mathbf{v}_{cmp}^{(in)}\cdot \mathbf{t}_3-\frac{m_p^2}{m_b}%
\mathbf{v}_{cmb}^{(in)}\cdot \mathbf{t}_i-m_p(\mathbf{w}_b^{(in)}\cdot
\mathbf{t}_1)(\mathbf{r}_{cmp}\cdot \mathbf{t}_2)\right\} =0.
\end{eqnarray}

After solving the equation for $x$, the solutions for the three remaining
quantities can be explicitly obtained in the forms
\begin{eqnarray}
x &=&-\frac{2\left( \mathbf{v}_{cmp}^{(in)}\cdot \mathbf{t}_3-\frac{m_p}{m_b}%
\mathbf{v}_{cmb}^{(in)}\cdot \mathbf{t}_i-(\mathbf{w}_b^{(in)}\cdot \mathbf{t%
}_1)(\mathbf{r}_{cmp}\cdot \mathbf{t}_2)\right) }{\left( \frac{m_p}2+\frac{%
m_p^3}{2m_b^2}+\frac I2(\frac{m_p}I)^2(\mathbf{r}_{cmp}\cdot \mathbf{t}%
_2)^2\right) }, \\
y &=&\frac{m_p}{m_b}\frac{2\left( \mathbf{v}_{cmp}^{(in)}\cdot \mathbf{t}_3-%
\frac{m_p}{m_b}\mathbf{v}_{cmb}^{(in)}\cdot \mathbf{t}_i-(\mathbf{w}%
_b^{(in)}\cdot \mathbf{t}_1)(\mathbf{r}_{cmp}\cdot \mathbf{t}_2)\right) }{%
\left( \frac{m_p}2+\frac{m_p^3}{2m_b^2}+\frac I2(\frac{m_p}I)^2(\mathbf{r}%
_{cmp}\cdot \mathbf{t}_2)^2\right) }, \\
z &=&\frac{m_p}I(\mathbf{r}_{cmp}\cdot \mathbf{t}_2)\frac{2\left( \mathbf{v}%
_{cmp}^{(in)}\cdot \mathbf{t}_3-\frac{m_p}{m_b}\mathbf{v}_{cmb}^{(in)}\cdot
\mathbf{t}_i-(\mathbf{w}_b^{(in)}\cdot \mathbf{t}_1)(\mathbf{r}_{cmp}\cdot
\mathbf{t}_2)\right) }{\left( \frac{m_p}2+\frac{m_p^3}{2m_b^2}+\frac I2(%
\frac{m_p}I)^2(\mathbf{r}_{cmp}\cdot \mathbf{t}_2)^2\right) }.
\end{eqnarray}

Finally, the searched final state quantities become expressed in terms of
the initial ones by the formulae
\begin{eqnarray}
\mathbf{v}_{cmb}^{(out)} &=&\mathbf{v}_{cmb}^{(in)}+x\text{ }\mathbf{t}_3,
\nonumber \\
\mathbf{v}_{cmp}^{(out)} &=&\mathbf{v}_{cmp}^{(in)}+y\text{ }\mathbf{t}_3,
\nonumber \\
\mathbf{w}_p^{(out)} &=&\mathbf{w}_p^{(in)},  \nonumber \\
\mathbf{w}_b^{(out)} &=&\mathbf{w}_b^{(in)}+z\text{ }\mathbf{t}_1,
\end{eqnarray}
which define the solution of the conservative shock problem. It
seems helpful to underline that for finding the given solution, we
have assumed energy conservation. However, even by considering the
frictionless case, can instead assume that a fraction of the total
mechanical energy could have been dissipated in other forms of
energy (vibrations, deformations, heat, etc.) during the impact. For
this purpose it only needed to added a dissipation term to the
energy conservation equation (\ref{conserv}).

\end{document}